\documentclass[lettersize,journal]{IEEEtran}
\usepackage{orcidlink}  
\usepackage{amsmath,amsfonts}
\usepackage{algorithmic}
\usepackage{algorithm}
\usepackage{array}
\usepackage{changes}
\usepackage[caption=false,font=normalsize,labelfont=sf,textfont=sf]{subfig}
\usepackage{textcomp}
\usepackage{stfloats}
\usepackage{url}
\usepackage{verbatim}
\usepackage{graphicx}
\usepackage{cite}
\usepackage{multirow}
\begin{document}

\newpage{© 20XX IEEE.  Personal use of this material is permitted.  Permission from IEEE must be obtained for all other uses, in any current or future media, including reprinting/republishing this material for advertising or promotional purposes, creating new collective works, for resale or redistribution to servers or lists, or reuse of any copyrighted component of this work in other works.}

\title{Array SAR 3D Sparse Imaging \\ Based on Regularization by Denoising \\ Under Few Observed Data}

\author{Yangyang Wang\textsuperscript{~\orcidlink{0000-0003-0095-603X}}\IEEEmembership{,~Member,~IEEE}, Xu Zhan\textsuperscript{~\orcidlink{0000-0003-28169791}}\IEEEmembership{,~Graduate~Student~Member,~IEEE},\\ Jing Gao, Jinjie Yao, Shunjun Wei and JianSheng Bai,
\thanks{This work was supported in part by the National Basic Research Program
	of China under Grant JCKY2021210B073 and Grant JCKY2022209A001; and in part by the Key Research and Development Plan of Shanxi Province under Grant 202102010101002. \textit{(Corresponding author: Xu Zhan).}}
\thanks{Y. Wang, J. Gao, J. Yao and J. Bai are with the School of information and Communication Engineering, North University of China, Taiyuan 030051, China.}
\thanks{X. Zhan and S. Wei are with the School of information and Communication Engineering, University of Electronic Science and Technology of China, Chengdu 611731, China (e-mail: zhanxu@std.uestc.edu.cn).}}

\markboth{Manuscript Accepted By IEEE TRANSACTIONS ON GEOSCIENCE AND REMOTE SENSING}%
{Shell \MakeLowercase{\textit{et al.}}: A Sample Article Using IEEEtran.cls for IEEE Journals}


\maketitle

\begin{abstract}
Array synthetic aperture radar (SAR) three-dimensional (3D) imaging can obtain 3D information of the target region, which is widely used in environmental monitoring and scattering information measurement. In recent years, with the development of compressed sensing (CS) theory, sparse signal processing is used in array SAR 3D imaging. Compared with matched filter (MF), sparse SAR imaging can effectively improve image quality. However, sparse imaging based on handcrafted regularization functions suffers from target information loss in few observed SAR data. Therefore, in this article, a general 3D sparse imaging framework based on Regulation by Denoising (RED) and proximal gradient descent type method for array SAR is presented. Firstly, we construct explicit prior terms via state-of-the-art denoising operators instead of regularization functions, which can improve the accuracy of sparse reconstruction and preserve the structure information of the target. Then, different proximal gradient descent type methods are presented, including a generalized alternating projection (GAP) and an alternating direction method of multiplier (ADMM), which is suitable for high-dimensional data processing. Additionally, the proposed method has robust convergence, which can achieve sparse reconstruction of 3D SAR in few observed SAR data. Extensive simulations and real data experiments are conducted to analyze the performance of the proposed method. The experimental results show that the proposed method has superior sparse reconstruction performance. 
\end{abstract}


\section{Introduction}
\IEEEPARstart{S}{ince} synthetic aperture radar (SAR) imaging is independent of weather conditions and light, which is widely used in military and civilian fields such as remote sensing \cite{ref1,ref2,ref3}, security inspection \cite{ref4}, and measurement of target scattering information \cite{ref5,ref6,refs1}. 3D SAR imaging can obtain the 3D scattering distribution of the observation region, which overcomes the limitations of traditional two-dimensional (2D) SAR imaging, such as shadow effects \cite{ref7}. In recent years, many different 3D SAR imaging mechanisms have been proposed, including interferometric SAR (InSAR) \cite{ref8}, topography SAR (TomoSAR) \cite{ref9}, and array SAR \cite{ref10,ref11,ref12}. InSAR estimates the elevation information of the scene by illuminating the target area from two different angles, which cannot directly reveal the 3D scattering distribution of the target region. TomoSAR synthesizes an aperture in the elevation direction by passing multiple times and collecting echo data, which can obtain the 3D scattering distribution of the target region. Compared to the above two mechanisms, array SAR has better adaptability and diverse working modes including side-looking, downward-looking, and forward-looking. Array SAR forms a 2D virtual array by performing linear motion in the azimuth and elevation directions ensuring the resolution of the 2D array. The range resolution is obtained by transmitting broadband signals.

3D SAR imaging methods can be mainly divided into two categories. The first type of SAR imaging method is based on the principle of matched filter (MF), which mainly includes time-domain and frequency-domain imaging methods \cite{ref13,ref14,ref15}. Among them, the classic time-domain imaging algorithm is the 3D back projection (BP) algorithm \cite{ref13}, which can achieve good imaging results, but its computational burden is unacceptable in real-time applications. The classic frequency-domain algorithm, the 3D wavenumber domain algorithm, can achieve imaging quickly, but the focus range is limited \cite{ref14}. The above imaging methods based on MF principle need to meet Nyquist–Shannon sampling theorem and imaging results of those methods are vulnerable to noise interference, which is difficult to meet the requirements of high-precision imaging. The second type of SAR imaging method is based on the sparse reconstruction principle, which can reconstruct target scattering information from incomplete measurement data.

SAR sparse reconstruction can be regarded as a typical inverse problem. The scattering information of the target region of interest can be reconstructed from measurement data using prior knowledge. The inverse problem of SAR sparse reconstruction is usually ill-posedness, which is difficult to directly the solution of the problem by only relying on a forward model representing the measurement system \cite{ref15}. Therefore, a prior model representing the prior knowledge of the target is required to alleviate the ill-posedness, which plays an important role in solving the inverse problem of SAR sparse reconstruction \cite{ref16}. In SAR sparse reconstruction, regularization functions are often used as prior models. In most sparse reconstruction methods, the selected regularization function is based on $L_1$ norm, which is an effective convex regularization function and can solve the inverse problem faster through iteration scheme \cite{ref17}. However, the $L_1$ regularization function is a biased estimation that often underestimates the scattering intensity of the target affecting the accuracy of reconstruction and subsequent image interpretation \cite{ref18,ref19,ref20}. Considering the above issues, another class of more complex nonconvex regularization functions has been proposed, including $L_q$ regularization function \cite{ref21,ref22}, smooth clipped absolute deviation (SCAD) \cite{ref23}, minimax convex penalty (MCP) \cite{ref24}, and Cauchy regularization function \cite{ref25}. Compared with $L_1$, nonconvex regularization functions can effectively reduce bias effects and improve reconstruction accuracy. 

The above nonconvex regularization functions have been widely used in 2D SAR imaging to improve imaging quality. However, the handcrafted regularization function may not fully represent target features in 3D SAR imaging affecting reconstruction accuracy and the optimization method based on regularization function requires to derive the explicit expression of the proximal operator when solving the imaging inverse problem, which is more difficult. To address the above issues, the Plug-and-Play (PnP) and Regularization by Denoising (RED) idea are considered to achieve 3D sparse imaging. Due to the use of implicit prior models, the definition of the objective function for PnP based imaging methods is often unclear, making it difficult to ensure robust convergence. In some application scenarios or data damage conditions (low sampling rate (SR) or signal-to-noise ratio (SNR)), it is necessary to use fewer observations to achieve 3D sparse imaging. However, existing regularization functions are difficult to characterize the multidimensional features of the target under few observed data, resulting in severe loss of target information and a subsequent decrease in image quality. Although PnP can improve image quality to a certain extent and preserve the feature information of the target, under the condition of few observed data, the resulting imaging may exhibit blurred target structures and artifacts, which affect subsequent image interpretation. Furthermore, ensuring the convergence of PnP is challenging, consequently diminishing the solution accuracy. Hence, in this article, we explore the construction of explicit prior terms through state-of-the-art denoisers for array SAR sparse imaging under RED. Firstly, SAR imaging with RED can more effectively preserve the structural features of targets under few observed SAR data conditions, while efficiently suppressing noise and artifacts interference. Secondly, RED formulates an explicit objective function for imaging optimization problems, which can better ensure convergence and imaging accuracy in the presence of few observed SAR data. Thirdly, RED exhibits strong applicability and can be employed across various imaging problems and scenarios.

In recent years, the proximal gradient descent type method has been widely used to solve the inverse problem of SAR imaging, which can be combined with RED for SAR imaging, such as the alternative direction method of multipliers (ADMM) and generalized alternating projection (GAP) \cite{refz1}. Therefore, in this article, we propose a novel array SAR 3D sparse imaging framework that combines RED and the proximal gradient descent type method. Compared with PnP, the proposed method constructs the prior through the denoising engine, which has the explicit objective function and robust convergence. Even under low SR conditions of the array SAR echo signal, the proposed method has superior sparse reconstruction performance. 
 
The main contributions of this article are summarized as follows.

1) A 3D sparse imaging framework for array SAR combining RED and proximal gradient descent type methods is proposed. The proposed framework has robust convergence and effectively preserve the structure information of the target even in few observed SAR data.
 
2) A prior term is construct through the denoising engine under the RED, which can better characterize the features of 3D SAR scenes under few observed data. To the best of our knowledge, this is the first exploration of RED to support 3D SAR imaging inverse problems. Additionally, the definition of the overall objective function is explicit and the convergence of the proposed method is easier to guarantee compare with PnP.

3) Different iterative algorithms of proximal gradient descent are presented to solve sparse SAR imaging inverse problems, including ADMM and GAP. The algorithm adopted in this article is suitable for high-dimensional imaging scenes. Additionally, different denoisers are used to construct the prior terms, including state-of-the-art deep denoisers. These indicate that the proposed imaging framework has flexibility and generality.

4) Extensive 3D SAR simulations and real data experiments are implemented to analyze the reconstruction performance of the proposed method. The results suggest that the proposed framework is effective and flexible for array SAR imaging.

The rest of this article is organized as follows. In Section II, a review of the related work is given. Section III introduces the proposed method in detail. The simulation and real data experimental results are reported in Section IV, along with performance analysis of the proposed method. In Section V, the characteristics of the proposed method are further discussed. Finally, the conclusion is presented in Section V.

\section{The Related Work}
\subsection{Sparse SAR Imaging}
{Traditional sparse SAR imaging methods are mostly based on regularization functions. At present, sparse imaging can be achieved through different optimization algorithm frameworks based on regularization functions, such as the approximate message passing (AMP) algorithm \cite{ref26}, the iterative shrinkage-thresholding algorithm (ISTA) \cite{ref27}, and the alternative direction method of multipliers (ADMM) algorithm \cite{ref28,ref29,ref30}. Xu \textit{et al}. proposed a 2D SAR sparse imaging method that combines nonconvex regularization functions and total variation norm, which significantly reduces bias effects and improves the reconstruction accuracy of the target \cite{ref3}. Karakus \textit{et al}. used Cauchy regularization function to solve the inverse problem of SAR imaging and sparse reconstruction performance of the proposed method is better than $L_1$ and the generalized minimax concave (GMC) penalty \cite{ref25}. Wu \textit{et al}. combined the complex approximated message passing (CAMP) method and $L_{1/2}$ regularization function to achieve 2D ISAR imaging, which enhances target features while suppressing scene noise \cite{ref21}. Bi \textit{et al}. proposed a sparse SAR imaging method based on $L_{2,1}$ regularization function, which can better reduce the ghost phenomenon compared to $L_1$ regularization function \cite{ref31}. However, in sparse imaging scenarios of array 3D SAR, most imaging methods are implemented using $L_1$ regularization functions. In our previous work, we first applied nonconvex regularization functions to sparse imaging of array 3D SAR, effectively improving the reconstruction accuracy, and achieved the measurement of 3D scattering information of the target \cite{ref6, ref20}.
	
In recent years, PnP and RED are considered to achieve sparse imaging in different fields. Venkatakrishnan \textit{et al}. first proposed a reconstruction framework based on PnP, which effectively improves the quality of magnetic resonance images (MRI) \cite{ref32}. Afterwards, some researchers show the superior performance of PnP for medical and optical image reconstruction. Chan \textit{et al}. used PnP for image super-resolution and analyzed the convergence and performance of the method \cite{ref33, ref34}. Sreehari \textit{et al}. proposed a sparse reconstruction method based on PnP to solve the problems of bright field electron tomography and sparse image interpolation \cite{ref35}. Additionally, a non-local means denoiser was further designed to ensure the convergence of the method. Teodoro \textit{et al}. combined scene adapted principles with PnP to ensure the convergence of the method to global minimum, using Gaussian mixture denoiser as a prior \cite{ref36}. Sun \textit{et al}. proposed a variant of PnP ADMM method that is suitable for large-scale settings \cite{ref37}. Wei \textit{et al}. proposed a tuning free PnP approximate method to solve a series of inverse imaging problems, which includes a policy network \cite{ref38}. Despite the aforementioned work \cite{ref39}, due to the use of implicit prior models, the definition of the objective function for imaging methods based on PnP is often unclear, making it difficult to ensure robust convergence. In some cases of few observed SAR data, the sparse reconstruction effect is not satisfactory. Therefore, in this article, we investigate the formulation of explicit prior using state-of-the-art denoisers under RED, which can better preserve the structure of the target and guarantee convergence. Romano \textit{et al}. first proposed RED for optical image reconstruction, which exhibits superior reconstruction performance \cite{ref40}. Reehorst \textit{et al}. combined RED with the gradient of the log-prior to accelerate convergence \cite{ref41}. RED shows great potential in solving imaging inverse problem. Although RED has shown superior potential in sparse reconstruction, research in the field of array SAR 3D imaging is still lacking.}

\subsection{The Sparse Imaging Model of Array SAR}
The geometric schematic diagram of array SAR imaging is shown in Fig. 1. X represents range direction, Y represents azimuth direction, and Z represents elevation direction. The system platform forms a virtual 2D array through linear motion in the azimuth and elevation directions obtaining array resolution. The transmission of broadband signals in the range direction provides guarantee for high resolution of the range direction \cite{ref42}. Therefore, the 3D spatial information of the target region is obtained. ${{\bf{P}}_e} = \left( {{x_e},{y_e},{z_e}} \right)$ represents the position of the array element and ${{\bf{P}}_t} = \left( {{x_t},{y_t},{z_t}} \right)$ represents the position of the scattering point $\delta $ in the imaging space. In this article, the transmission signal of the array SAR system is linear frequency modulated (LFM). The expression of the echo signal received by the system sensor is as follows

\begin{figure}[!t]
	\centering
	\includegraphics[width=3.5in]{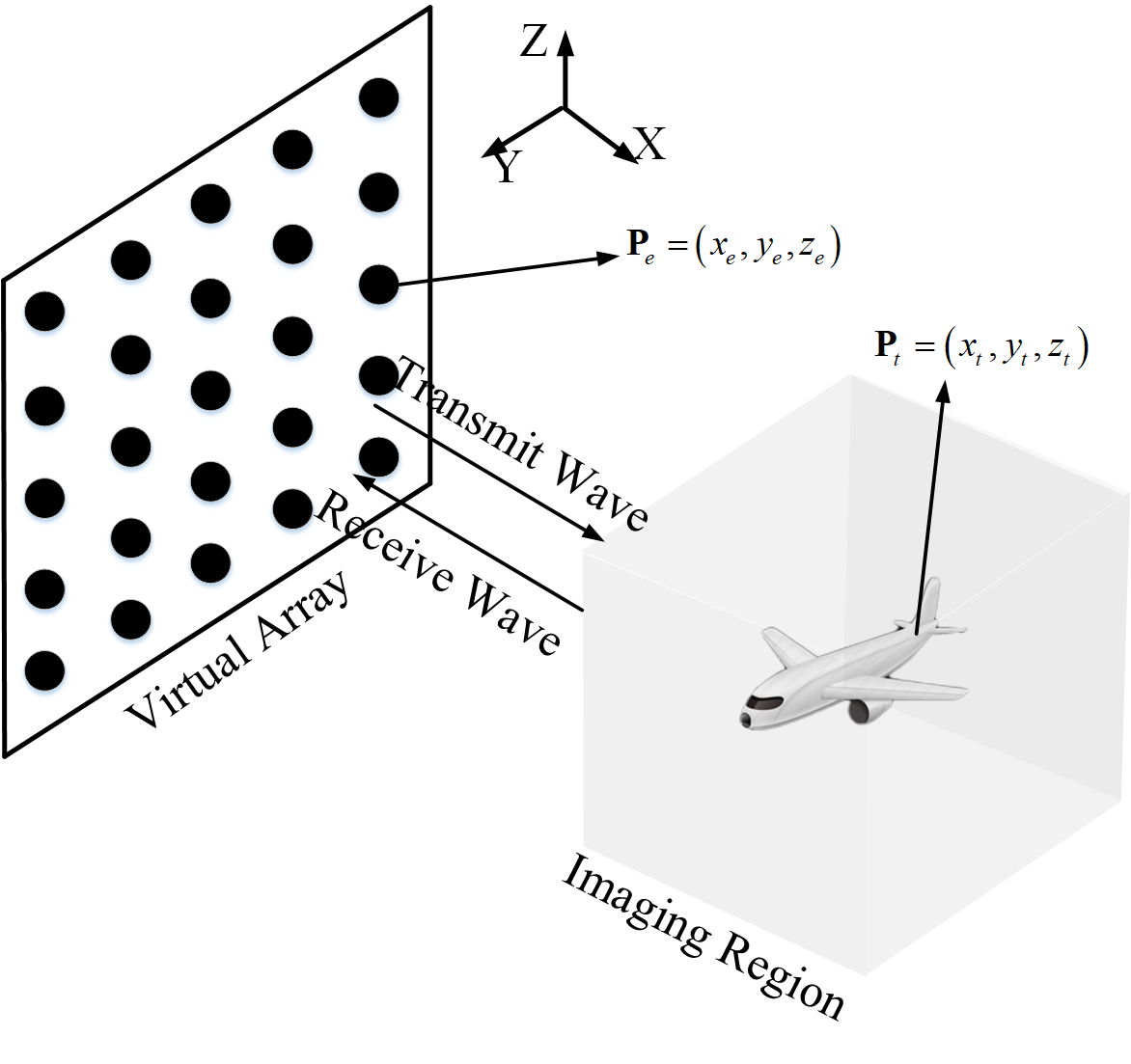}
	\caption{The array SAR measurement configuration.}
	\label{fig_1}
\end{figure}

\begin{equation}
	\label{Eq1}
	{\bf{s}}\left( {{{\bf{P}}_e},t} \right) = \int\limits_{{{\bf{P}}_t} \in \Omega } {\sigma \left( {{{\bf{P}}_t}} \right)\exp \left\{ { - j\pi K{{\left( {t - \frac{R}{c}} \right)}^2} - j2kR} \right\}d{{\bf{P}}_t}} 
\end{equation}
where $t$ is the range fast time and $K$ is the frequency modulation slope of the LFM signal. $k = {{2\pi {f_c}} \mathord{\left/
		{\vphantom {{2\pi {f_c}} c}} \right.
		\kern-\nulldelimiterspace} c}
$ is the wavenumber, $c$ is the speed of light, and ${f_c}$ is the carrier frequency. $R = {\left\| {{{\bf{P}}_e} - {{\bf{P}}_t}} \right\|_2}$ represents distance history and $\Omega $ represents imaging space.

The echo of the r-th distance unit after pulse compression is

\begin{equation}
	\label{Eq2}
	{{\bf{s}}_c}\left( {{{\bf{P}}_e},r} \right) = \int\limits_{{{\bf{P}}_t} \in {\Omega _r}} {\sigma \left( {{{\bf{P}}_t}} \right)\chi \left( r \right)\exp \left\{ { - j2kR} \right\}d{{\bf{P}}_t}} 
\end{equation}
where $\chi \left( r \right)$ is the range ambiguity function. To express concisely, let $\beta \left( {{{\bf{P}}_t}} \right) = \sigma \left( {{{\bf{P}}_t}} \right)\chi \left( r \right)$, then the discretization form of \eqref{Eq2} is

\begin{equation}
	\label{Eq3}
	{{\bf{s}}_c}\left( {{{\bf{P}}_e},r} \right) = \sum\limits_{{{\bf{P}}_t} \in {\Omega _r}} {\beta \left( {{{\bf{P}}_t}} \right)\exp \left\{ { - j2kR} \right\}}  
\end{equation}

Let M be the total number of array elements, the model in \eqref{Eq3} can be written as a linear model

\begin{equation}
	\label{Eq4}
	{\bf{y}} = {\bf{Ax}}  
\end{equation}
Considering the impact of noise in the actual environment, the sparse reconstruction model of array SAR is

\begin{equation}
	\label{Eq5}
	{\bf{y}} = {\bf{Ax}} + {{\bf{n}}_\varepsilon } 
\end{equation}
where ${\bf{y}} \in {C^{M \times 1}}$ is the measurement data, ${\bf{x}} \in {C^{N \times 1}}$ is the complex scattering information of the imaging scene, ${{\bf{n}}_\varepsilon } \in {C^{M \times 1}}$ is the noise vector, and ${\bf{A}} \in {C^{M \times N}}$ is the system measurement matrix.

Reconstruction of the scattering information of the imaging scene from the measured data belongs to a typical inverse problem \cite{ref43, ref44, refs2}, which can be solved by the appropriate optimization algorithm, which will be formulated in the next section.

\section{Array SAR 3D Sparse Imaging Based on Regularization by Denoising and Proximal Gradient Descent Type Method}
Firstly, the typical ADMM framework for solving the inverse problem of SAR imaging is briefly introduced, along with some defects. {Then, given the existing problems, the array SAR 3D sparse imaging framework based on RED and proximal gradient descent type method is formulated in detail. In this article, two different proximal gradient descent methods, ADMM and GAP, are presented.}

\subsection{The Typical ADMM for Sparse Imaging}
Following Bayesian theory, the inverse problem of SAR imaging is equivalent to the maximum a posteriori probability (MAP) \cite{ref45}. Thus, the optimization problem of sparse reconstruction can be expressed in the following form

\begin{equation}
	\label{Eq6}
	\begin{aligned}
		{\bf{\hat x}} &= \mathop {\arg \max }\limits_{\bf{x}} P\left( {{\bf{x}}|{\bf{y}}} \right) = \mathop {\arg \max }\limits_{\bf{x}} {{P\left( {{\bf{y}}|{\bf{x}}} \right)P\left( {\bf{x}} \right)} \mathord{\left/
				{\vphantom {{P\left( {{\bf{y}}|{\bf{x}}} \right)P\left( {\bf{x}} \right)} {P\left( {\bf{y}} \right)}}} \right.
				\kern-\nulldelimiterspace} {P\left( {\bf{y}} \right)}}\\
		&= \mathop {\arg \min }\limits_{\bf{x}} f\left( {\bf{x}} \right) + \lambda g\left( {\bf{x}} \right)
	\end{aligned}
\end{equation}
where $f\left( {\bf{x}} \right) =  - \log P\left( {{\bf{y}}|{\bf{x}}} \right)$, $\lambda g\left( {\bf{x}} \right) =  - \log P\left( {\bf{x}} \right)$. By utilizing the monotonic increasing property of logarithmic functions, the maximization problem in \eqref{Eq6} is transformed into a minimization problem, which belongs to a typical unconstrained optimization problem. The data term $f\left( {\bf{x}} \right)$ ensures that the reconstructed scattering information of imaging scene is consistent with the echo data \cite{ref46}, expressed as

\begin{equation}
	\label{Eq7}
	f\left( {\bf{x}} \right) = \frac{1}{{2{\sigma ^2}}}\left\| {{\bf{y}} - {\bf{Ax}}} \right\|_2^2
\end{equation}
When ${\sigma ^2}$ is absorbed into the regularization parameter $\lambda $, the data term is $f\left( {\bf{x}} \right) = {{\left\| {{\bf{y}} - {\bf{Ax}}} \right\|_2^2} \mathord{\left/
		{\vphantom {{\left\| {{\bf{y}} - {\bf{Ax}}} \right\|_2^2} 2}} \right.
		\kern-\nulldelimiterspace} 2}
$. $\lambda $ is a regular parameter. $g\left( {\bf{x}} \right)$ represents a prior term. Regularization functions are usually used as a prior to alleviates the ill-posedness of the sparse reconstruction problem. The commonly regularization functions used in SAR sparse reconstruction mainly include convex regularization functions, such as $L_1$, and nonconvex regularization functions, such as MCP. The $L_1$ regularization function is ${\left\| {\bf{x}} \right\|_1}$ and the proximal operator ${\tau _{{L_1}}}$ of $L_1$ is as follows

\begin{equation}
	\label{Eq8}
	{\tau _{{L_1}}}\left( {{\bf{y}}_S^{(i)};\lambda } \right) = \left\{ {\begin{array}{*{20}{l}}
			{{\bf{y}}_S^{(i)} - \lambda ,}&{{\bf{y}}_S^{(i)} \ge \lambda }\\
			{0,}&{\left| {{\bf{y}}_S^{(i)}} \right| < \lambda }\\
			{{\bf{y}}_S^{(i)} + \lambda ,}&{{\bf{y}}_S^{(i)} \le  - \lambda }
	\end{array}} \right.
\end{equation}

The definition of MCP and the corresponding proximal operator ${\tau _{{MCP}}}$ is as follows

\begin{equation}
	\label{Eq9}
	\begin{aligned}
	{g_{MCP}}\left( {\bf{x}} \right) = \sum\limits_{j = 1}^N & {\left( {\left| {{x_j}} \right| - \frac{{{{\left| {{x_j}} \right|}^2}}}{{2\theta }}} \right)}  \odot 1\left( {\left| {{x_j}} \right| \le \theta \lambda } \right) \\
	&+ \frac{\theta }{2} \odot 1\left( {\left| {{x_j}} \right| > \theta \lambda } \right)
    \end{aligned}
\end{equation}

\begin{equation}
	\label{Eq10}
	{\tau _{MCP}}\left( {{\bf{y}}_S^{(i)};\lambda } \right) = \left\{ {\begin{array}{*{20}{l}}
			{0,}&{\left| {{\bf{y}}_S^{(i)}} \right| < \lambda }\\
			{\frac{{\theta \left( {\left| {{\bf{y}}_S^{(i)}} \right| - \lambda } \right)sign\left( {{\bf{y}}_S^{(i)}} \right)}}{{\theta  - 1}},}&{\lambda  \le \left| {{\bf{y}}_S^{(i)}} \right| \le \theta \lambda }\\
			{{\bf{y}}_S^{(i)},}&{\left| {{\bf{y}}_S^{(i)}} \right| > \theta \lambda }
	\end{array}} \right.	
\end{equation}
where $ \odot $ denotes element-wise multiplication and $\theta  > 1$.   

\eqref{Eq6} is an unconstrained optimization problem that can be transformed into a more easily solved constrained optimization problem using variable splitting \cite{ref47}. The basic concept of variable splitting is to create a new variable ${\bf{v}}$ and impose constraints ${\bf{v}} = {\bf{x}}$ on $g\left( {\bf{x}} \right)$, resulting in the following constrained optimization problems

\begin{equation}
	\label{Eq11}
	\begin{array}{l}
		\mathop {\min }\limits_{\bf{x}} f\left( {\bf{x}} \right) + \lambda g\left( {\bf{v}} \right)\\
		{\rm{subject \ to \ }}{\bf{v}} = {\bf{x}}
	\end{array}	
\end{equation}

\begin{figure}[!t]
	\centering
	\includegraphics[width=3.5in]{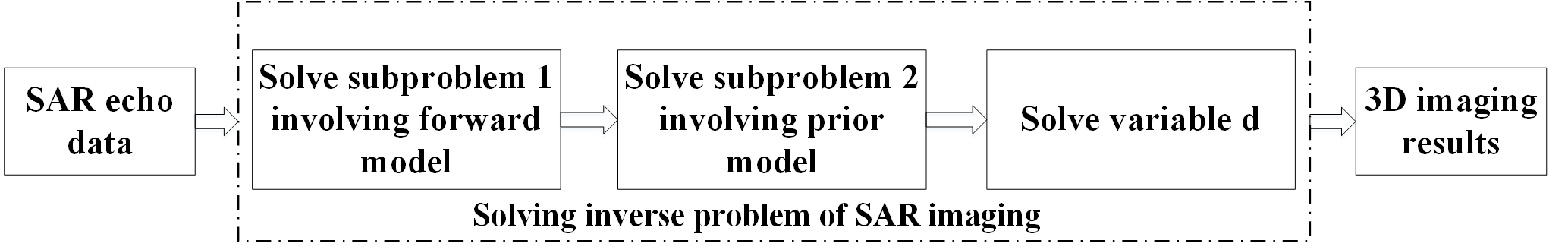}
	\caption{Basic block diagram of SAR imaging with ADMM.}
	\label{fig_2}
\end{figure}

Next, we utilize the method of multipliers (MEM), namely the augmented Lagrangian method (ALM) \cite{ref48}, to define the augmented Lagrangian function of optimization problem (11)

\begin{equation}
	\label{Eq12}
	L\left( {{\bf{x}},{\bf{v}},\mu } \right) = f\left( {\bf{x}} \right) + \lambda g\left( {\bf{v}} \right) + \frac{\mu }{2}\left\| {{\bf{x}} - {\bf{v}} - {\bf{d}}} \right\|_2^2
\end{equation}
where $\mu $ is the penalty parameter. Solving multiple variables simultaneously increases the difficulty of optimization problems. Therefore, ADMM is adopted to solve one of the variables alternately by fixing the remaining variables. The basic block diagram of solving SAR imaging inverse problem based on ADMM is shown in Fig. 2. The original optimization problem is transformed into solving multiple subproblems and the expression corresponding to the basic module of ADMM is as follows

\begin{equation}
	\label{Eq13}
	{\rm{subproblem1}}:{{\bf{x}}_t} = \mathop {\arg \min }\limits_{\bf{x}} \frac{\mu }{2}\left\| {{\bf{x}} - {{\bf{v}}_{t - 1}} - {{\bf{d}}_{t - 1}}} \right\|_2^2 + f\left( {\bf{x}} \right)
\end{equation}

\begin{equation}
	\label{Eq14}
	{\rm{subproblem2}}:{{\bf{v}}_t} = \mathop {\arg \min }\limits_{\bf{v}} \frac{\mu }{2}\left\| {{{\bf{x}}_t} - {\bf{v}} - {{\bf{d}}_{t - 1}}} \right\|_2^2 + \lambda g\left( {\bf{v}} \right)
\end{equation}

\begin{equation}
	\label{Eq15}
	{{\bf{d}}_t} = {{\bf{d}}_{t - 1}} - {{\bf{x}}_t} + {{\bf{v}}_t}
\end{equation}

In the process of solving subproblem 1, ${\bf{x}}$ is solved by fixing variables ${\bf{v}}$ and ${\bf{d}}$. Therefore, $g\left( {\bf{v}} \right)$ in \eqref{Eq12} can be ignored. The solution of \eqref{Eq13} is equivalent to minimizing a strict convex quadratic function, usually solved by the optimization method, i.e., least square method.

{For subproblem 2, the regularization function is used to characterize the prior information of the objective and to solve the subproblem using proximal operators. However, sparse imaging method based on regularization function seriously lose target information and reduce imaging performance under SAR data damage conditions, i.e., low SR or low SNR.}

\subsection{Array SAR 3D Sparse Imaging Based on Regularization by Denoising and ADMM}
{To achieve high-quality 3D SAR imaging under few observed data, a sparse imaging method based on RED and ADMM (RADMM) is presented. For the subproblem involving forward imaging models, the solving process is as the following method}

\begin{equation}
	\label{Eq16}
	{{\bf{x}}_t} = {\left( {{{\bf{A}}^H}{\bf{A}} + \mu {\bf{I}}} \right)^{ - 1}}\left( {{{\bf{A}}^H}{\bf{y}} + \mu {{\bf{v}}_{t - 1}} + \mu {{\bf{d}}_{t - 1}}} \right)
\end{equation}

For subproblem 2, ${\bf{v}}$ is solved through fixed variables ${\bf{x}}$ and ${\bf{d}}$. In normal SAR sparse imaging, the handcrafted regularization is used as a prior term and the corresponding proximal operator is used to solve subproblem 2 to improve SAR image quality. \eqref{Eq8} and \eqref{Eq10} present two different proximal operators. Among them, \eqref{Eq8} is the proximal operator of $L_1$, and \eqref{Eq10} is the proximal operator of MCP. Compared with imaging methods based on MF, sparse imaging utilizes regularization functions as a prior to ensure the enhancement of target features while suppressing sidelobes and noise. At present, the most commonly used regularization function in 3D SAR imaging is the convex function $L_1$, which is a biased estimation \cite{ref49}. Compared to $L_1$, nonconvex regularization functions can better improve SAR image quality. In our previous work, we combined nonconvex regularization functions with majorization-minimization (MM) to achieve 3D SAR imaging \cite{ref12}. However, the handcrafted regularization function may have insufficient feature representation in 3D SAR imaging under few observed SAR data, which limits the accuracy of reconstruction. Although PnP can preserve target features to some extent, it exhibits blurred target structures and artifacts under few observed SAR data.

To address the above issues, we construct an explicit prior ${g_R}\left( {\bf{v}} \right)$ based on RED to better preserve the features of the target \cite{ref40}.

\begin{equation}
	\label{Eq20}
	{g_R}\left( {\bf{v}} \right) = \frac{1}{2}{{\bf{v}}^H}\left( {{\bf{v}} - D\left( {\bf{v}} \right)} \right)
\end{equation}
where $D$ is the state-of-the-art denoising operator. Subproblem 2 becomes the following form

\begin{equation}
	\label{Eq21}
	{{\bf{v}}_t} = \mathop {\arg \min }\limits_{\bf{v}} \frac{\mu }{2}\left\| {{{\bf{x}}_t} - {\bf{v}} - {{\bf{d}}_{t - 1}}} \right\|_2^2 + \frac{\lambda }{2}{{\bf{v}}^H}\left( {{\bf{v}} - D\left( {\bf{v}} \right)} \right)
\end{equation}

By making the gradient of \eqref{Eq21} zero, we obtain the following relationship

\begin{equation}
	\label{Eq22}
	\mu \left( {{\bf{v}} - {{\bf{x}}_t} + {{\bf{d}}_{t - 1}}} \right) + \lambda \left( {{\bf{v}} - D\left( {\bf{v}} \right)} \right) = 0
\end{equation}
According to the fixed-point strategy, ${\bf{v}}$ can be solved iteratively as follows

\begin{equation}
	\label{Eq23}
	{{\bf{v}}_t} = \frac{\lambda }{{\lambda  + \mu }}D\left( {{{\bf{v}}_{t - 1}}} \right) + \frac{\mu }{{\lambda  + \mu }}\left( {{{\bf{x}}_t} - {{\bf{d}}_{t - 1}}} \right)
\end{equation}

Compared with the current solving method based on handcrafted regularization functions, the proposed method does not require the derivation of proximal operators and better preserve the structural details of the target. Compared with the imaging method based on PnP, the proposed method has explicit prior terms and objective functions, which is easy to guarantee robust convergence and is suitable for few observed SAR data conditions. The basic step of the proposed method is shown in Algorithm 1.

\begin{algorithm}[!t]
	\caption{The basic procedure of RADMM.}\label{Alg1}
	\begin{algorithmic}
		\STATE 
		\STATE {\textbf{Input:}} 
		\STATE \hspace{0.2cm} Maximum number of iterations ${t_{\max }}$; Error parameter $\varpi $;  ${\bf{y}}$; ${{\bf{v}}_0}$; ${{\bf{d}}_0}$.
		\STATE
		\STATE \hspace{0.2cm} \textbf{While} $1 \le t \le {t_{\max }}$ and ${\mathop{\rm Re}\nolimits}  > \varpi $ \textbf{do}
		\STATE \hspace{0.8cm} \textbf{Solve subproblem 1}:
		\STATE \hspace{1.6cm} ${{\bf{x}}_t} = {\left( {{{\bf{A}}^H}{\bf{A}} + \mu {\bf{I}}} \right)^{ - 1}}\left( {{{\bf{A}}^H}{\bf{y}} + \mu {{\bf{v}}_{t - 1}} + \mu {{\bf{d}}_{t - 1}}} \right)
		$
		\STATE \hspace{0.8cm} \textbf{Solve subproblem 2}:
		\STATE \hspace{1.6cm} Let ${{\bf{m}}_0} = {{\bf{v}}_{t - 1}}$, ${\bf{\hat m}} = {{\bf{x}}_t} - {{\bf{d}}_{t - 1}}$
		\STATE \hspace{1.6cm} \textbf{For} $j = {\rm{1,2}}...{\rm{,J}}$ \textbf{do}
		\STATE \hspace{2.4cm} ${\bf{\bar m}} = D\left( {{{\bf{m}}_{j - 1}}} \right)$
		\STATE \hspace{2.4cm} ${{\bf{m}}_j} = \frac{\lambda }{{\lambda  + \mu }}{\bf{\bar m}} + \frac{\mu }{{\lambda  + \mu }}{\bf{\hat m}}$
		\STATE \hspace{1.6cm} \textbf{End}
		\STATE \hspace{1.6cm} ${{\bf{v}}_t} = {{\bf{m}}_{\rm{J}}}$
		\STATE \hspace{0.8cm} \textbf{Solve} ${{\bf{d}}_t}$:
		\STATE \hspace{1.6cm} ${{\bf{d}}_t} = {{\bf{d}}_{t - 1}} - {{\bf{x}}_t} + {{\bf{v}}_t}$
		\STATE \hspace{0.8cm} \textbf{Update the residual:}
		\STATE \hspace{0.8cm} ${\mathop{\rm Re}\nolimits}  = {{\left\| {{{\bf{x}}_t} - {{\bf{x}}_{t - 1}}} \right\|} \mathord{\left/
				{\vphantom {{\left\| {{{\bf{x}}_t} - {{\bf{x}}_{t - 1}}} \right\|} {\left\| {{{\bf{x}}_{t - 1}}} \right\|}}} \right.
				\kern-\nulldelimiterspace} {\left\| {{{\bf{x}}_{t - 1}}} \right\|}}
		$
		\STATE \hspace{0.8cm} $ t=t+1 $
		\STATE \hspace{0.2cm} \textbf{End While}
		\STATE 
		\STATE {\textbf{Output:}}
		\STATE \hspace{0.2cm} Sparse imaging results ${{\bf{x}}_t}$
	\end{algorithmic}
\end{algorithm}

\subsection{Array SAR 3D Sparse Imaging Based on Regularization by Denoising and GAP (RGAP)}
To verify the generality of the proposed framework, GAP \cite{refz1} is incorporated into the framework to achieve 3D sparse imaging of array SAR. Unlike the optimization problem \eqref{Eq11} of ADMM, GAP achieves sparse imaging by solving the following optimization problem

\begin{equation}
	\label{Eq24}
	\begin{array}{l}
		\left( {\widehat {\bf{x}},\widehat {\bf{v}}} \right) = \mathop {\arg \min }\limits_{{\bf{x}},{\bf{v}}} \frac{1}{2}\left\| {{\bf{x}} - {\bf{v}}} \right\|_2^2 + \lambda g\left( {\bf{v}} \right)\\
		{\rm{subject \ to }} \ {\bf{y}} = {\bf{Ax}}
	\end{array}
\end{equation}

In order to simplify the solution, \eqref{Eq24} is decoupled into two subproblem. Firstly, it is to solve for the variable ${\bf{x}}$. ${{\bf{x}}_t}$ achieves iterative updates through the Euclidean projection of ${{\bf{v}}_{t - 1}}$ on ${\bf{y}} = {\bf{Ax}}$ 

\begin{equation}
	\label{Eq25}
	{{\bf{x}}_t} = {{\bf{v}}_{t - 1}} + {{\bf{A}}^H}{\left( {{\bf{A}}{{\bf{A}}^H}} \right)^{ - 1}}\left( {{\bf{y}} - {\bf{A}}{{\bf{v}}_{t - 1}}} \right)
\end{equation}

Afterwards, based on the RED concept, the variable ${\bf{v}}$ can also be solved by constructing a prior term using the state-of-art denoiser. Based on the fixed point strategy, ${\bf{v}}$ can be iteratively solved

\begin{equation}
	\label{Eq26}
	{{\bf{v}}_t} = \frac{1}{{\lambda  + 1}}{\bf{x}} - \frac{\lambda }{{\lambda  + 1}}D\left( {{{\bf{v}}_{t - 1}}} \right)
\end{equation}

Next, we will analyze the differences between ADMM and GAP. In theory, in the case of noise free, the performance of ADMM and GAP is similar \cite{refz2}. In the noise case, due to the consideration of noise in the ADMM model, its sparse imaging performance is usually better than GAP. From another perspective, the constraint imposed by GAP is ${\bf{y}} = {\bf{A}}\widehat {\bf{x}}$, which ignores the noise present in most scenarios. In the presence of noise, the solution obtained by GAP may deviate from the true value. Unlike GAP, ADMM solves ${\bf{x}}$ by minimizing $\left\| {{\bf{y}} - {\bf{Ax}}} \right\|_2^2$, rather than based on the aforementioned constraints. Therefore, under noisy conditions, the solution of ADMM is closer to the true value compared to GAP. This indicates the generality of the proposed framework, which can be integrated with different iterative algorithms to achieve high-quality 3D sparse SAR imaging.

Finally, we analyze the convergence of the proposed framework in array SAR imaging. The optimization problem of RED can be reformulated via Rockafellar function \cite{refz3}. We provide some basic definitions based on variational analysis \cite{refz3, refz4, refzz1}.

\textit{Definition 1}: For each group of points $\left\{ {{x_1},{x_2},...,{x_n}} \right\}$ and any $n \ge 2$, if the continuous mapping $B$ satisfies \eqref{Eq27}, then $B$ is called the maximum cyclically monotone.

\begin{equation}
	\label{Eq27}
	\sum\limits_{j = 1}^n {\left( {B\left( {{x_j}} \right),{x_{j + 1}} - {x_j}} \right)}  \le 0,{\rm{where \ }}{x_1} = {x_{j + 1}}
\end{equation}

\textit{Definition 2}: For each group of points $\left\{ {{x_1},{x_2},...,{x_n}} \right\}$ and any $n \ge 2$, if $D$ satisfies \eqref{Eq28}, then $D$ is called cyclically firm nonexpansive.

\begin{equation}
	\label{Eq28}
	\sum\limits_{j = 1}^n {\left( {{x_j} - D\left( {{x_j}} \right),D\left( {{x_j}} \right) - D\left( {{x_{j + 1}}} \right)} \right)}  \ge 0,{\rm{where \ }}{x_1} = {x_{j + 1}}
\end{equation}
Afterwards, by Proposition 1 in \cite{refz4}, the maximum cyclically monotone and cyclically firmly nonexponential are linked, which can help analyze the convergence of sparse imaging frameworks based on RED.

\textit{Proposition 1}: $B = {I_d} - D$ denotes displacement mapping and $D$ is cyclically firmly nonexpansive. Then $B$ satisfies maximal cyclically monotone.

Based on the above definition and proposition, the Rockafellar function is defined via regularization.

\textit{Definition 3}: Let $u$ be any point, for $n \ge 2$, define the function ${F_1}\left( {x;u} \right)$ with parameter $u$

\begin{equation}
	\label{Eq29}
{F_1}\left( {x;u} \right) \buildrel \Delta \over = \left\{ {\begin{array}{*{20}{c}}
		{\left( {x,B\left( u \right)} \right) - \left( {u,A\left( u \right)} \right),}&{n = 2}\\
		{\sup \left( \begin{array}{l}
				\left( {x - {x_{j - 1}},B\left( {{x_{j - 1}}} \right)} \right)\\
				+ \sum\limits_{j = 1}^{n - 2} {\left( {{x_{j + 1}} - {x_j},B\left( {{x_j}} \right)} \right),} 
			\end{array} \right)}&{otherwise}
\end{array}} \right.
\end{equation}
Hence, the Rockafterlar function can be defined as
\begin{equation}
	\label{Eq30}
	{F_R}\left( {x;u} \right) \buildrel \Delta \over = \mathop {\sup }\limits_{n \in \left( {2,3,...} \right)} {F_n}\left( {x;u} \right)
\end{equation}

After that, Proposition 2 is given \cite{refz5}, which indicates the advantageous characteristics of the Rockafellar function.

\textit{Proposition 2}: If $B$ is maximum cyclically monotone, then ${F_R}\left( {x;u} \right)$ is convex, lower semicontinuous. For any $x$, there is the following relationship
\begin{equation}
	\label{Eq31}
	B\left( x \right) = \nabla {F_R}\left( {x;u} \right)
\end{equation}

Considering the above propositions, we can express the imaging inverse problem as the following optimization form
\begin{equation}
	\label{Eq32}
	\mathop {\min }\limits_{\bf{x}} {F_{Rock}}\left( {\bf{x}} \right) = f\left( {\bf{x}} \right) + \lambda {F_R}\left( {x;u} \right)
\end{equation}
where $D$ is continuous denoiser. For sparse SAR imaging, $f\left( {\bf{x}} \right)$ is ${{\left\| {{\bf{y}} - {\bf{Ax}}} \right\|_2^2} \mathord{\left/
		{\vphantom {{\left\| {{\bf{y}} - {\bf{Ax}}} \right\|_2^2} {2{\sigma ^2}}}} \right.
		\kern-\nulldelimiterspace} {2{\sigma ^2}}}
$ where ${\bf{y}}$ is the echo data and ${\bf{A}}$ is the system measurement matrix. The denoiser residual is defined as $B = {I_d} - D$. ${I_d}$ denotes the identity operator. The objective function of the sparse optimization problem based on RED is
\begin{equation}
	\label{Eq33}
	{F_{RED}} = \mathop {\arg \min }\limits_{\bf{x}} f\left( {\bf{x}} \right) + \frac{\lambda }{2}\left( {{\bf{x}},{\bf{x}} - D\left( {\bf{x}} \right)} \right)
\end{equation}
Next, Theorem 1 is given, which presents convergence conditions of the sparse imaging based on RED. 

\textit{Theorem 1}: If $D$ is cyclically firmly nonexponential, then \eqref{Eq32} is a convex optimization problem, and the gradient of its objective function is $\nabla {F_{RED}} = \nabla f\left( {\bf{x}} \right) + \lambda \left( {{\bf{x}} - D\left( {\bf{x}} \right)} \right)$.

\textit{Proof}: Combining Proposition 1 and Proposition 2, residual $B = {I_d} - D$ is maximal cyclically monotone. Therefore, the function ${F_R}\left( {x;u} \right)$ is convex differentiable, and its gradient is $\nabla {F_R}\left( {x;u} \right) = B\left( {\bf{x}} \right) = {\bf{x}} - D\left( {\bf{x}} \right)$, which indicates the following relationship
\begin{equation}
	\label{Eq34}
	\begin{aligned}
		\nabla {F_{Rock}}\left( {\bf{x}} \right) &= \nabla f\left( {\bf{x}} \right) + \lambda \nabla {F_R}\left( {x;u} \right)\\
		&= \nabla f\left( {\bf{x}} \right) + \lambda \left( {{\bf{x}} - D\left( {\bf{x}} \right)} \right)\\
		&= \nabla {F_{RED}}\left( {\bf{x}} \right)
	\end{aligned}
\end{equation}
The conditions for minimizing convex optimization problem \eqref{Eq32} through RED can be obtained, thus verifying the convergence of the sparse imaging method based on RED. The sufficient condition for the convergence of the sparse imaging method based on RED is that $D$ is the maximum cyclically monotone, which allows the denoiser to be non-differentiable.

\section{Experiments and Analysis}

\begin{figure*}[!t]
	\centering
	\subfloat[]{\includegraphics[width=1.3in]{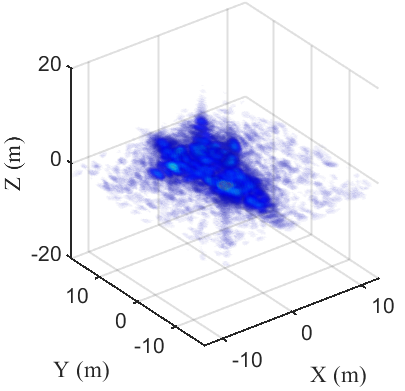}%
		\label{fig3_a}}
	\hfil
	\subfloat[]{\includegraphics[width=1.3in]{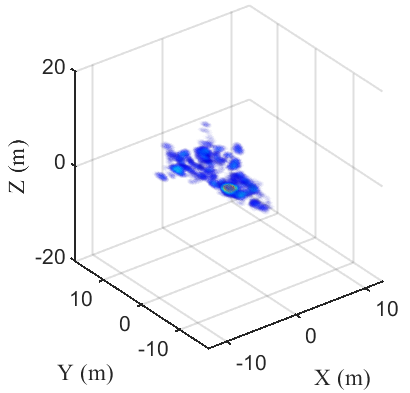}%
		\label{fig3_b}}
	\hfil
	\subfloat[]{\includegraphics[width=1.3in]{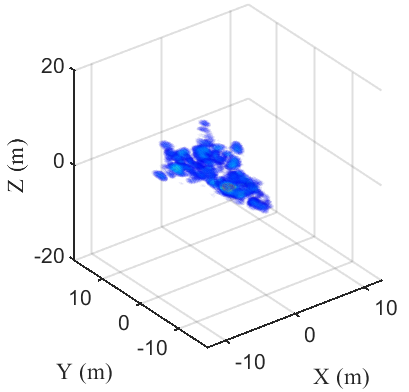}%
		\label{fig3_c}}
	\hfil
	\subfloat[]{\includegraphics[width=1.3in]{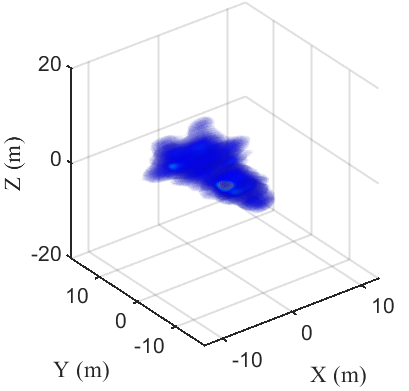}%
		\label{fig3_d}}
	\hfil
	\subfloat[]{\includegraphics[width=1.3in]{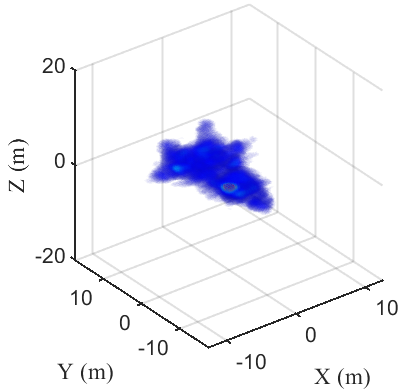}%
		\label{fig3_e}}
	\hfil
	\subfloat[]{\includegraphics[width=1.3in]{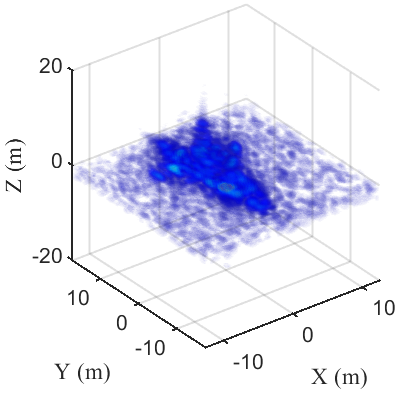}%
		\label{fig3_f}}
	\hfil
	\subfloat[]{\includegraphics[width=1.3in]{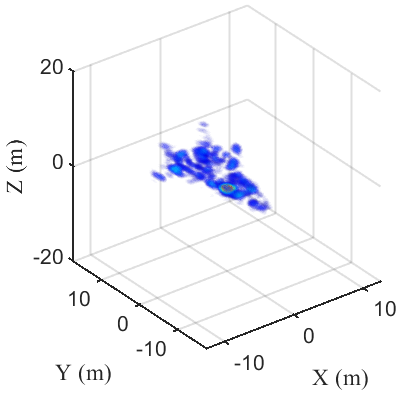}%
		\label{fig3_g}}
	\hfil
	\subfloat[]{\includegraphics[width=1.3in]{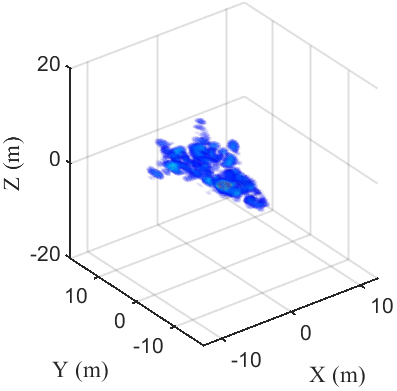}%
		\label{fig3_h}}
	\hfil
	\subfloat[]{\includegraphics[width=1.3in]{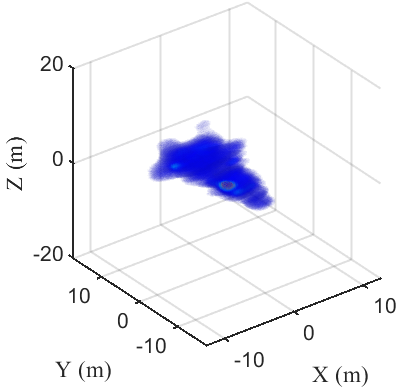}%
		\label{fig3_i}}
	\hfil
	\subfloat[]{\includegraphics[width=1.3in]{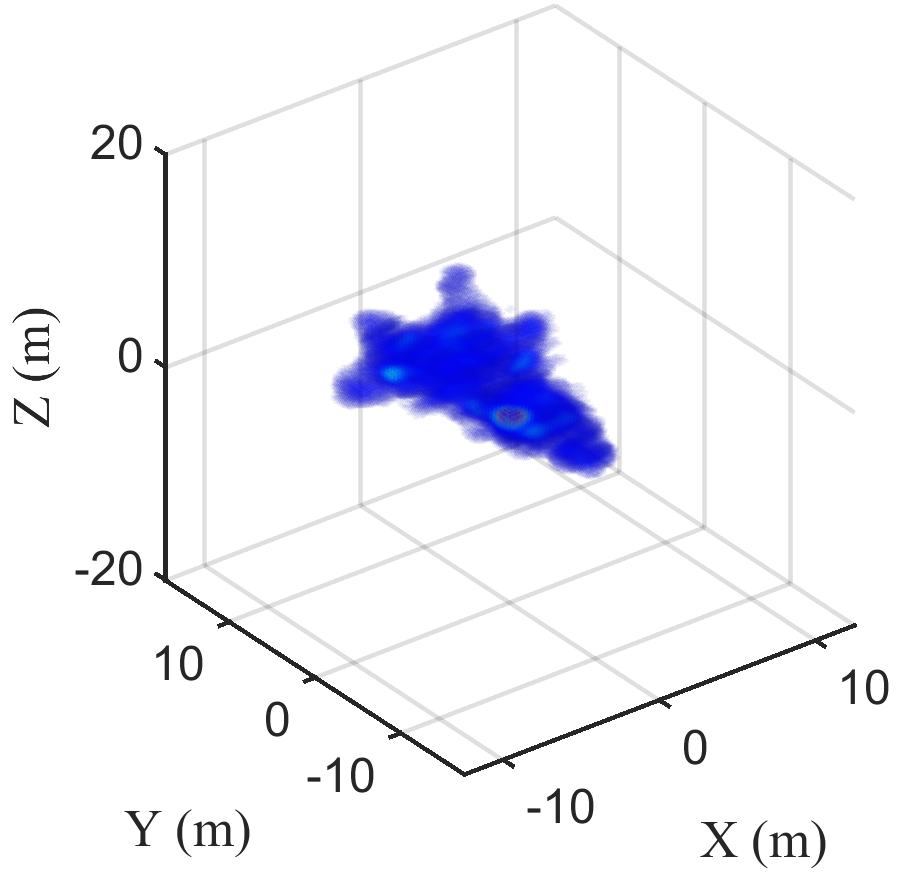}%
		\label{fig3_j}}
	\hfil
	\subfloat[]{\includegraphics[width=1.3in]{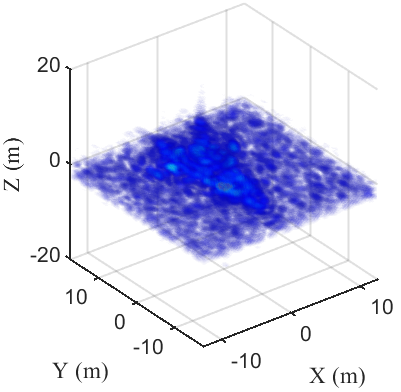}%
		\label{fig3_k}}
	\hfil
	\subfloat[]{\includegraphics[width=1.3in]{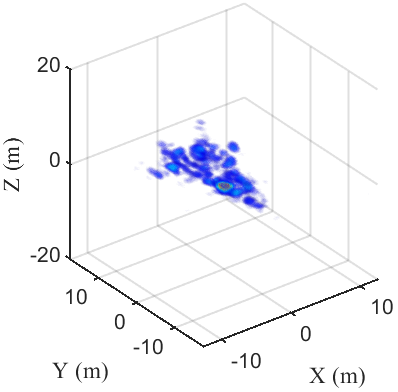}%
		\label{fig3_l}}
	\hfil
	\subfloat[]{\includegraphics[width=1.3in]{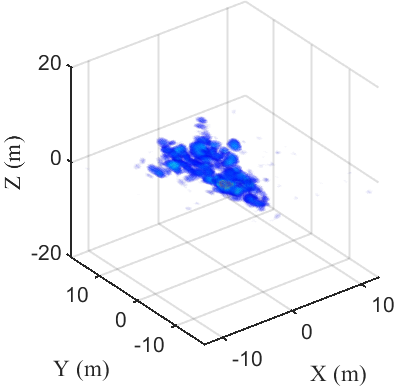}%
		\label{fig3_m}}
	\hfil
	\subfloat[]{\includegraphics[width=1.3in]{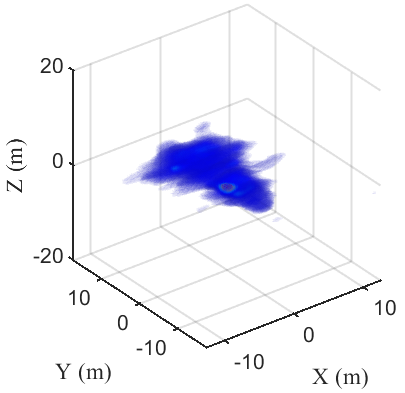}%
		\label{fig3_n}}
	\hfil
	\subfloat[]{\includegraphics[width=1.3in]{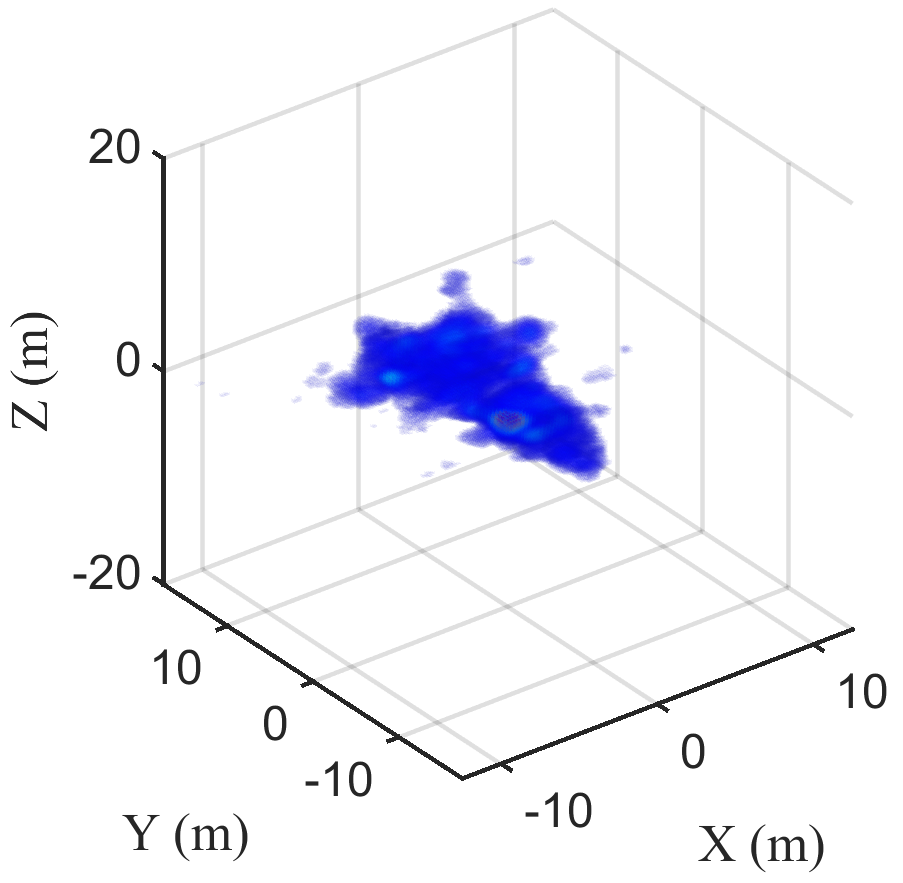}%
		\label{fig3_o}}
	\hfil
	\subfloat[]{\includegraphics[width=1.3in]{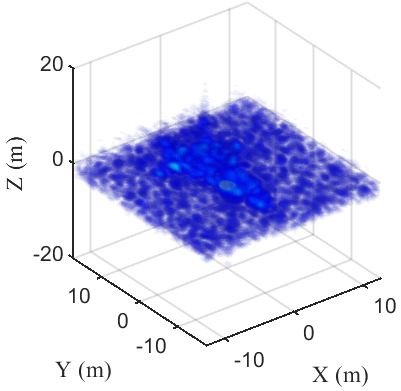}%
		\label{fig3_p}}
	\hfil
	\subfloat[]{\includegraphics[width=1.3in]{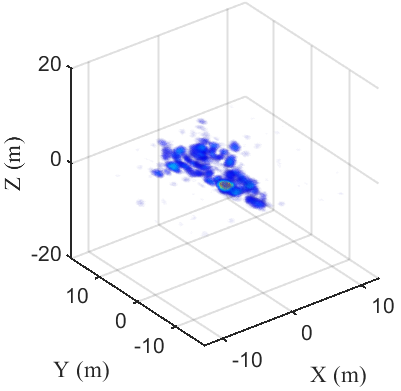}%
		\label{fig3_q}}
	\hfil
	\subfloat[]{\includegraphics[width=1.3in]{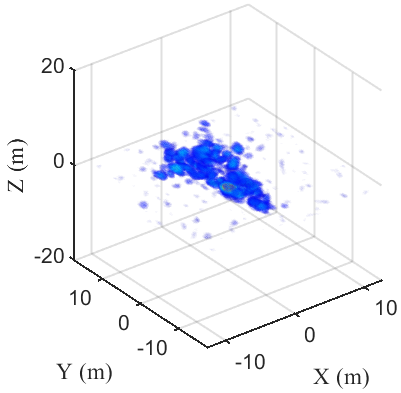}%
		\label{fig3_r}}
	\hfil
	\subfloat[]{\includegraphics[width=1.3in]{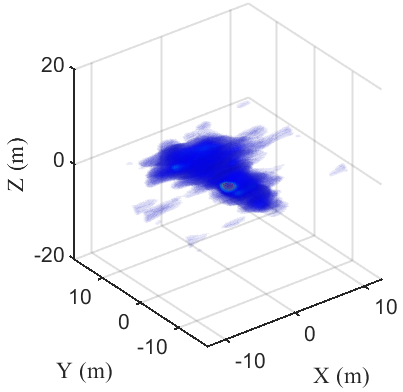}%
		\label{fig3_s}}
	\hfil
	\subfloat[]{\includegraphics[width=1.3in]{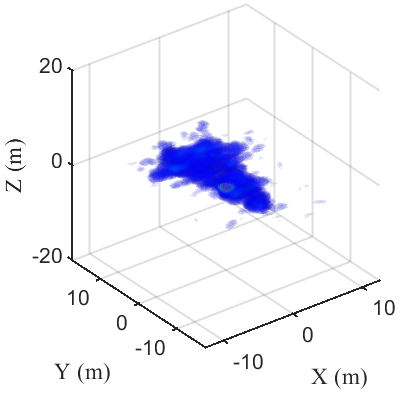}%
		\label{fig3_t}}
	\caption{The imaging results of the airplane under different SR. (a)-(e) The result of MF, $L_1$, GMCP, PnP and the proposed method at 75$\%$ SR. {(f)-(j) The result of MF, $L_1$, GMCP, PnP and the proposed method at 50$\%$ SR. (k)-(o) The result of MF, $L_1$, GMCP, PnP and the proposed method at 25$\%$ SR. (p)-(t) The result of MF, $L_1$, GMCP, PnP and the proposed method at 15$\%$ SR.}}
	\label{fig3}
\end{figure*} 

In this section, different simulation and real data experiments are carried out to verify the effectiveness of the proposed method. Firstly, we presented the simulation experiment to analyze the performance of the method under different SR and SNR. Then, we presented three different sets of SAR experiments to further validate the effectiveness of the proposed method in different experimental scenarios. The denoising operator used in the experiment is the nonlocal means (NLM). Three indicators, peak signal-to-noise ratio (PSNR), structural similarity index for measuring (SSIM) \cite{ref52} and normalized mean square error (NMSE), are used to evaluate the performance of the methods. We compare the proposed method with MF, $L_1$, GMCP and PnP. Visualization and quantitative analysis are presented to further analyze the sparse reconstruction performance of the proposed method. SSIM is defined as follows

\begin{equation}
	\label{Eq35}
	{\rm{SSIM}} = \frac{{\left( {2E\left( {\bf{x}} \right)E\left( {{\bf{\hat x}}} \right) + {L_1}} \right)\left( {2{\sigma _{{\bf{x\hat x}}}} + {L_2}} \right)}}{{\left( {{E^2}\left( {\bf{x}} \right) + {E^2}\left( {{\bf{\hat x}}} \right) + {L_1}} \right)\left( {\sigma _{\bf{x}}^2 + \sigma _{{\bf{\hat x}}}^2 + {L_2}} \right)}}
\end{equation}
where ${L_1} = {\left( {{k_1}L} \right)^2}$, ${L_2} = {\left( {{k_2}L} \right)^2}$. $k_1$ and $k_2$ are set to 0.01 and 0.03 by default. $E\left( {\bf{x}} \right)$ and ${\sigma _{\bf{x}}}$ denote the mean value and standard deviation of image ${\bf{x}}$, respectively. $E\left( {{\bf{\hat x}}} \right)$ and ${\sigma _{{\bf{\hat x}}}}$ denote the mean value and standard deviation of ${\bf{\hat x}}$, respectively. $L$ represents the dynamic range of the pixel-values. The larger the SSIM, the better the performance of the method.

\subsection{Aircraft Simulation}

\begin{figure*}[!t]
	\centering
	\subfloat[]{\includegraphics[width=1.3in]{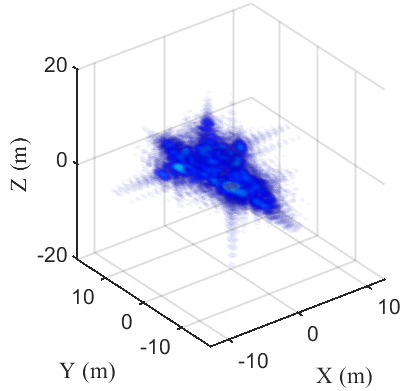}%
		\label{fig4_a}}
	\hfil
	\subfloat[]{\includegraphics[width=1.3in]{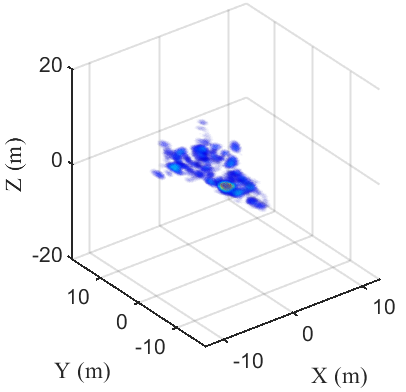}%
		\label{fig4_b}}
	\hfil
	\subfloat[]{\includegraphics[width=1.3in]{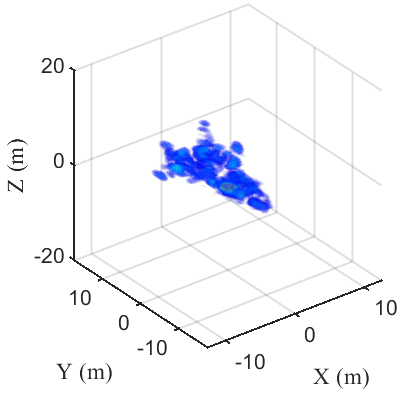}%
		\label{fig4_d}}
	\hfil
	\subfloat[]{\includegraphics[width=1.3in]{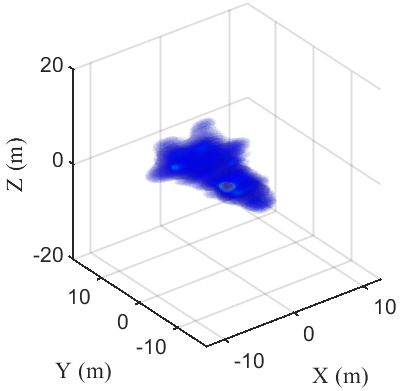}%
		\label{fig4_e}}
	\hfil
	\subfloat[]{\includegraphics[width=1.3in]{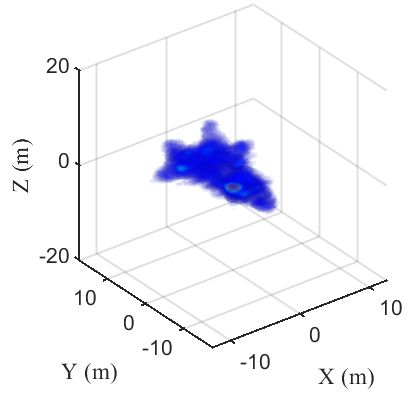}%
		\label{fig4_f}}
	\hfil
	\subfloat[]{\includegraphics[width=1.3in]{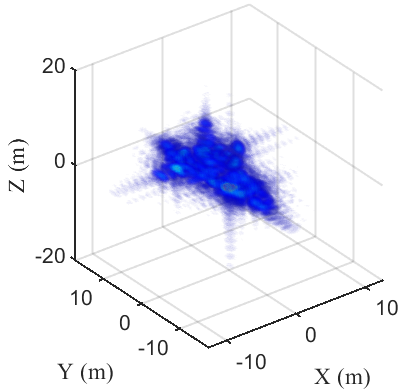}%
		\label{fig4_g}}
	\hfil
	\subfloat[]{\includegraphics[width=1.3in]{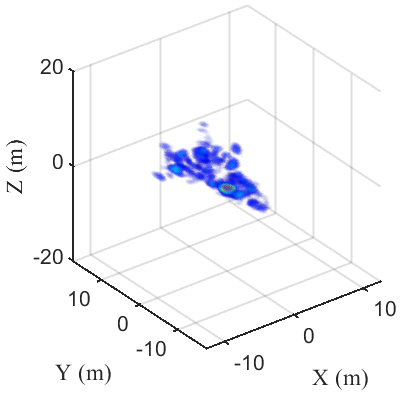}%
		\label{fig4_h}}
	\hfil
	\subfloat[]{\includegraphics[width=1.3in]{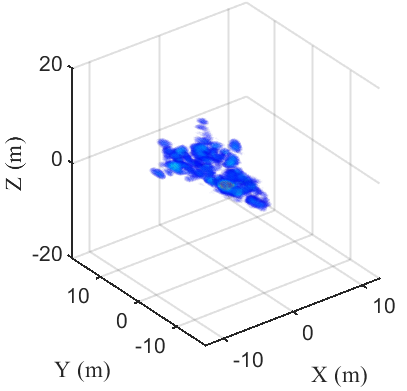}%
		\label{fig4_j}}
	\hfil
	\subfloat[]{\includegraphics[width=1.3in]{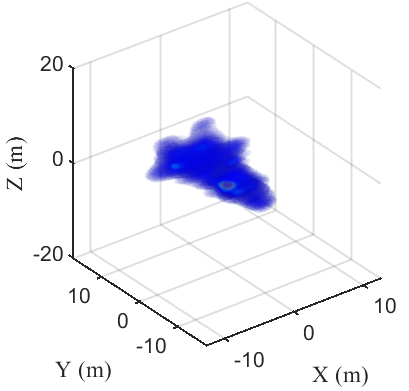}%
		\label{fig4_k}}
	\hfil
	\subfloat[]{\includegraphics[width=1.3in]{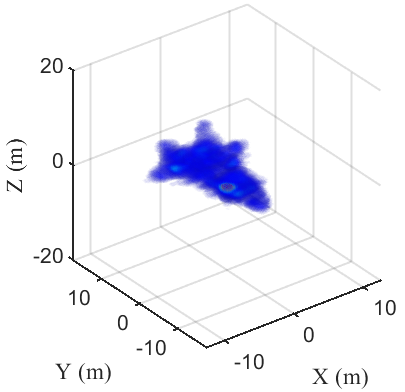}%
		\label{fig4_l}}
	\caption{The imaging results of the airplane at different SNR. (a)-(f) The result of MF, $L_1$, GMCP, PnP and the proposed method at 18 dB SNR. (g)-(l) The result of MF, $L_1$, GMCP, PnP and the proposed method at 12 dB SNR.}
	\label{fig4}
\end{figure*} 

\begin{figure}[!t]
	\centering
	\includegraphics[width=3.5in]{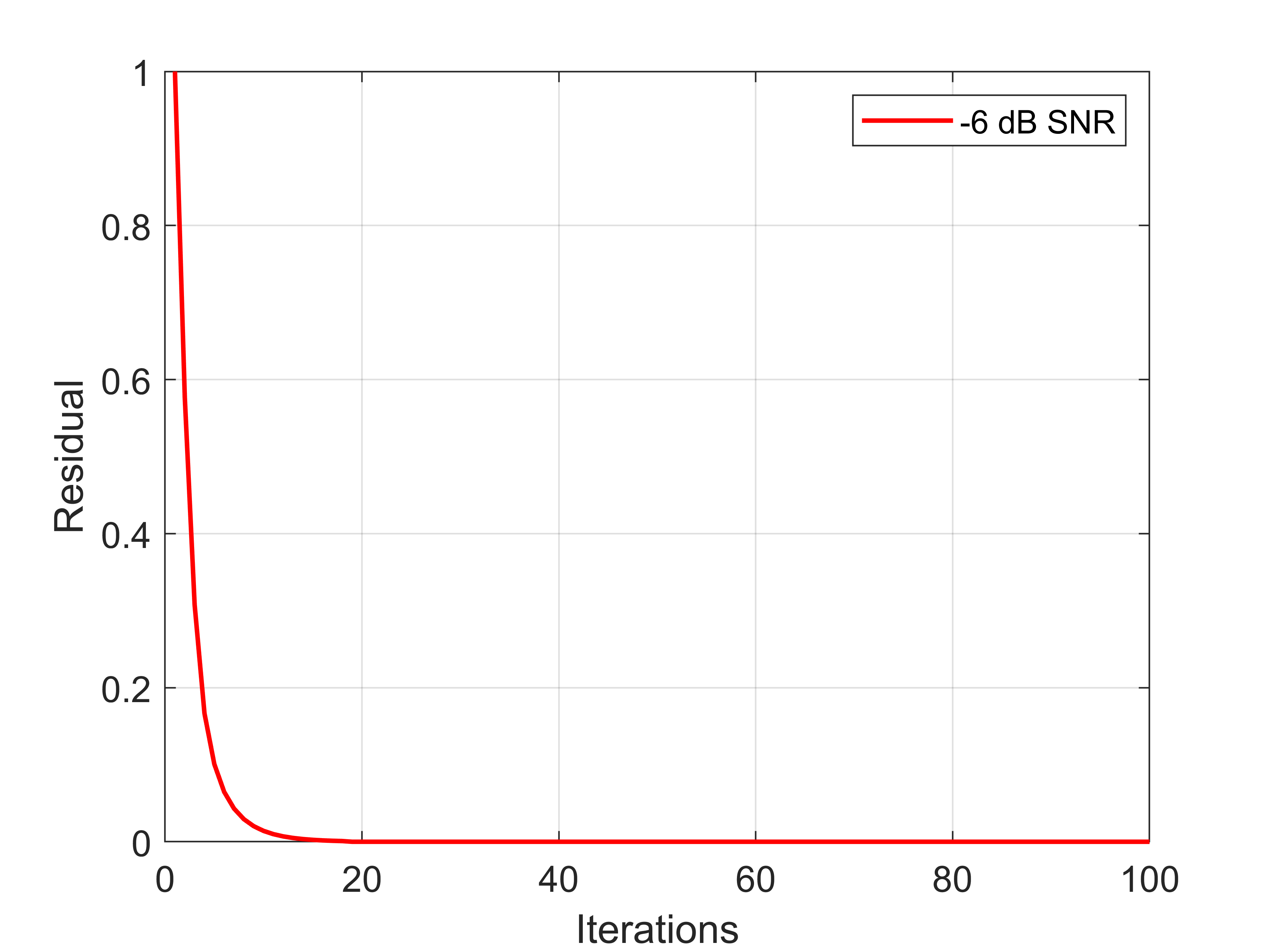}
	\caption{The convergence curve.}
	\label{fig_5}
\end{figure}

Firstly, a 3D simulation experiment of the aircraft model is carried out to verify the effectiveness of the proposed method. The main simulation parameters are as follows. The carrier frequency is 37.5 GHz and the bandwidth is 0.164 GHz. The size of the planar array is 3 m × 3 m. The 3D imaging results at 75$\%$, 50$\%$ and 25$\%$ SR are shown in Fig. 3.

Fig. 3(a)-(e) shows the 3D reconstruction results of MF, $L_1$, GMCP, PnP and the proposed method at 75$\%$ SR. {Fig. 3(f)-(j) shows the 3D reconstruction results of MF, $L_1$, GMCP, PnP and the proposed method at 50$\%$ SR. Fig. 3(k)-(o) shows the 3D reconstruction results of MF, $L_1$, GMCP, PnP and the proposed method at 25$\%$ SR. Fig. 3(p)-(t) shows the 3D reconstruction results of MF, $L_1$, GMCP, PnP and the proposed method at 15$\%$ SR.} The visualization results show that the proposed method can effectively improve the quality of SAR images and better preserve the detailed information of contour structures. The indicators with different sampling rates are listed in Table~\ref{tab1}. {For 75$\%$ sampling, the PSNR of $L_1$, GMCP, PnP and the proposed method are 40.9109 dB, 43.9291 dB, 47.7338 dB and 49.8614 dB, respectively. It can be seen that among these methods, the PSNR of the proposed method is the highest. The SSIM of $L_1$, GMCP, PnP and the proposed method are 0.8807, 0.9184, 0.9765 and 0.9870, respectively. For 50$\%$ sampling, among the five different methods, the PSNR and SSIM of the proposed method are still the highest, with 48.2375 dB and 0.9760, respectively. The NMSE of the proposed method is the lowest, 0.1164. For 25$\%$ and 15$\%$ sampling, the indicator of the proposed method is better than other methods.} The results show that although the performance of the sparse reconstruction method decreases as the sampling rate decreases, that of the proposed method is still superior to $L_1$, GMCP, PnP.

Next, we analyze the performance of the proposed method {under different SNR conditions of the array SAR echo signal}. Fig. 4 shows the 3D sparse reconstruction results of $L_1$, GMCP, PnP and the proposed method at SNR of 18 dB and 12 dB. The results show that the proposed method significantly improves the reconstruction accuracy compared to $L_1$, GMCP and PnP. The quantitative analysis is listed in Table~\ref{tab2}. As the SNR decreases, the PSNR and SSIM of four methods decrease and the NMSE increases accordingly. Nevertheless, the sparse reconstruction performance of the proposed method is still superior to the other methods even at lower SNR.

Fig. 5 shows the convergence curve of the proposed method at SNR of -6 dB, which demonstrates that the proposed method still exhibits robust convergence at low SNR.

\begin{table*}[!t]
	\caption{PSNR, SSIM and NMSE of aircraft model under different sampling\label{tab1}}
	\centering
	\begin{tabular}{|c|c|c|c|c|c|c|c|c|c|c|c|c|}
		\hline
		\multirow{2}{*}{SR} & \multicolumn{3}{c|}{$L_1$} & \multicolumn{3}{c|}{GMCP} & \multicolumn{3}{c|}{PnP} & \multicolumn{3}{c|}{RADMM}\\  
		\cline{2-13}
		& PSNR & SSIM & NMSE & PSNR & SSIM & NMSE & PSNR & SSIM & NMSE & PSNR & SSIM & NMSE \\
		\hline
		75 & 40.9109 & 0.8807 & 0.4069 & 43.9291 & 0.9184 & 0.2416 & 47.7338 & 0.9765 & 0.1556 & 49.8614 & 0.9870 & 0.0914\\
		\hline
		{50} & {40.2792} & {0.8509} & {0.4440} & {42.9006} & {0.8899} & {0.2889} &  {45.8420} &  {0.9291} & {0.3461} & {48.2375} & {0.9760} & {0.1164}\\
		\hline
		{25} & {38.4065} & {0.7663} & {0.5436} & {40.4402} &  {0.8118} & {0.4059} & {44.4588}  & {0.9010} & {0.3801} & {47.9923} & {0.9680} & {0.1237}\\
		\hline
		{15} & {36.6844} & {0.7292} & {0.5725} & {38.6247} &  {0.7792} & {0.4366} & {43.7934}  & {0.8883} & {0.3956} & {46.9360} & {0.9279} & {0.1627}\\
		\hline
	\end{tabular}
\end{table*}

\begin{table*}[!t]
	\caption{PSNR, SSIM and NMSE of aircraft model under different SNR\label{tab2}}
	\centering
	\begin{tabular}{|c|c|c|c|c|c|c|c|c|c|c|c|c|}
		\hline
		\multirow{2}{*}{SNR} & \multicolumn{3}{c|}{$L_1$} & \multicolumn{3}{c|}{GMCP} & \multicolumn{3}{c|}{PnP} & \multicolumn{3}{c|}{RADMM}\\  
		\cline{2-13}
		& PSNR & SSIM & NMSE & PSNR & SSIM & NMSE & PSNR & SSIM & NMSE & PSNR & SSIM & NMSE\\
		\hline
		18 & 41.3404 & 0.9001 & {0.3863} & 44.6631 & 0.9363 & {0.2141} & 48.0208 & 0.9783 & {0.1527} & 50.2811 & 0.9882 & {0.0870}\\
		\hline
		12 & 41.3237 & 0.8868 & {0.3893} & 44.5411 & 0.9231 & {0.2189} & 47.9817 & 0.9778 & {0.1533} & 50.2177 & 0.9877 & {0.0878} \\
		\hline
		6 & 40.9673 & 0.8465 & {0.4061} & 44.0816 & 0.8867 & {0.2362} & 47.8094 & 0.9759 & {0.1553} & 49.9773 & 0.9863 & {0.0904} \\
		\hline
		0 & 40.2038 & 0.7482 & {0.4461} & 42.6497 & 0.7888 & {0.3010} & 47.2151 & 0.9697 & {0.1631} & 49.5013 & 0.9853 & {0.0924} \\
		\hline
		-6 & 37.7527 & 0.5363 & {0.5961} & 39.4779 & 0.5768 & {0.4785} & 45.1673 & 0.9299 & {0.2107} & 48.0436  &  0.9797 & {0.1676} \\
		\hline
	\end{tabular}
\end{table*}

\begin{figure*}[!t]
	\centering
	\subfloat[]{\includegraphics[width=1.5in]{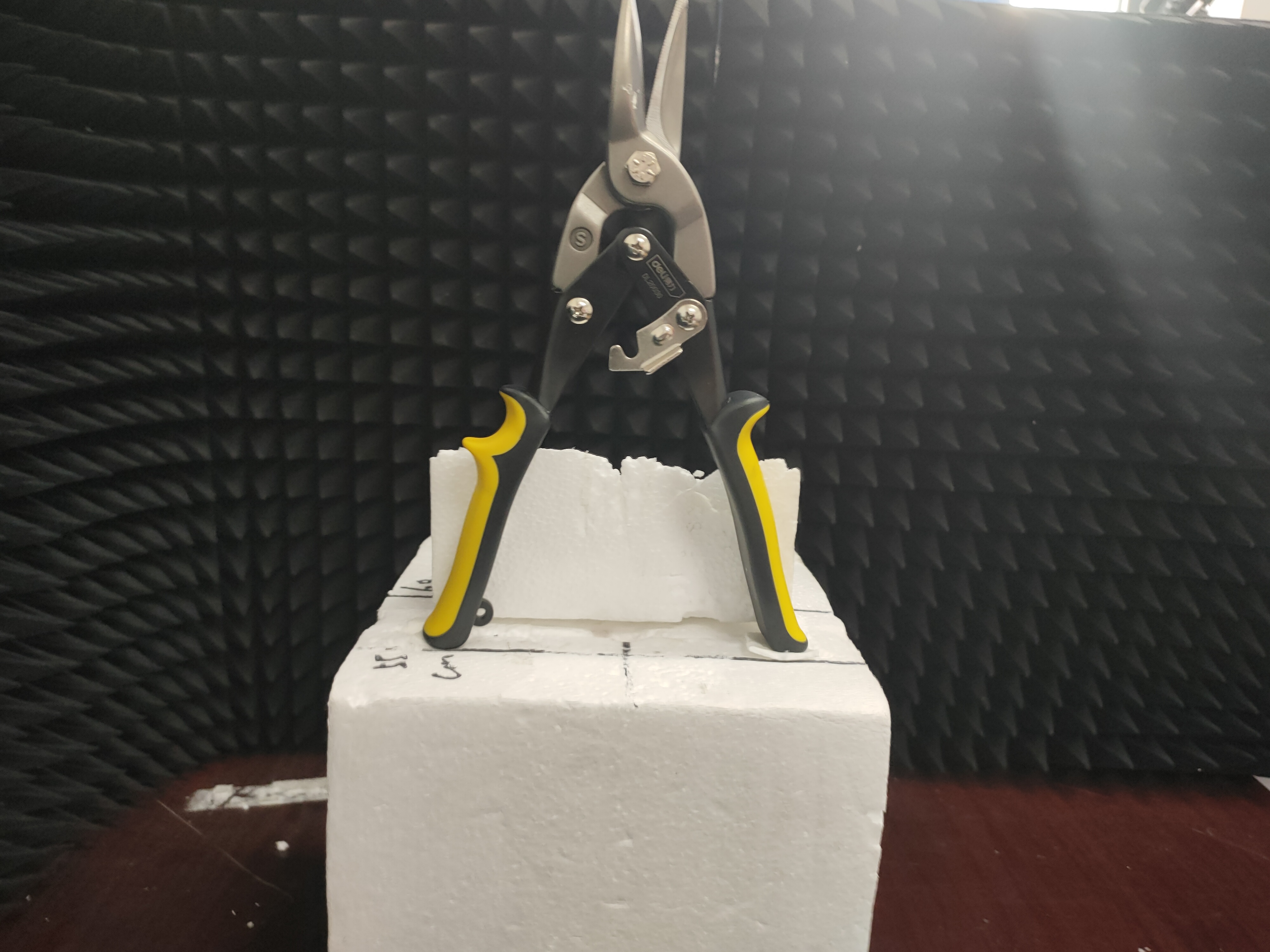}%
		\label{fig7_a}}
	\hfil
	\subfloat[]{\includegraphics[width=1.5in]{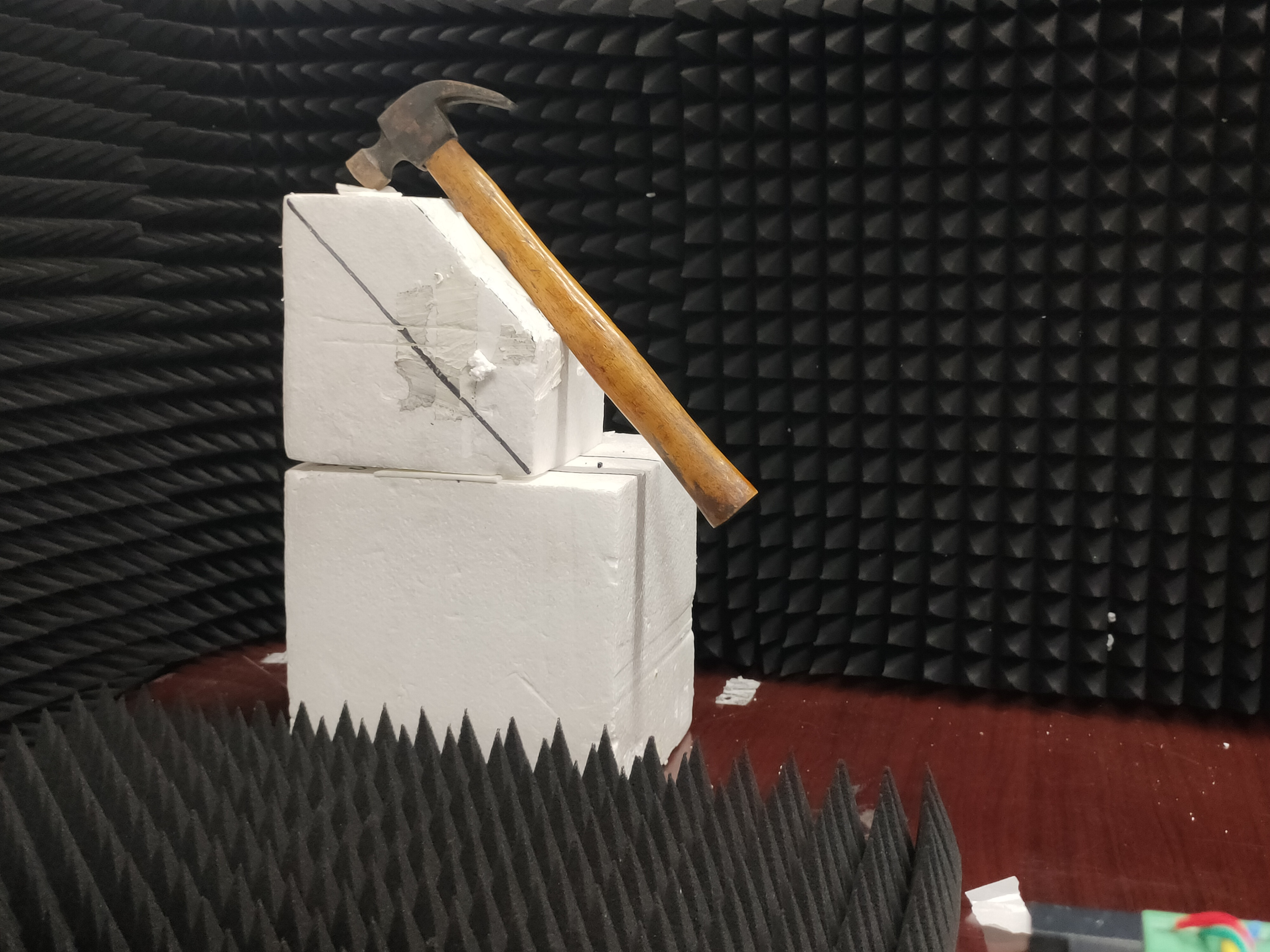}%
		\label{fig7_b}}
	\hfil
	\subfloat[]{\includegraphics[width=1.5in]{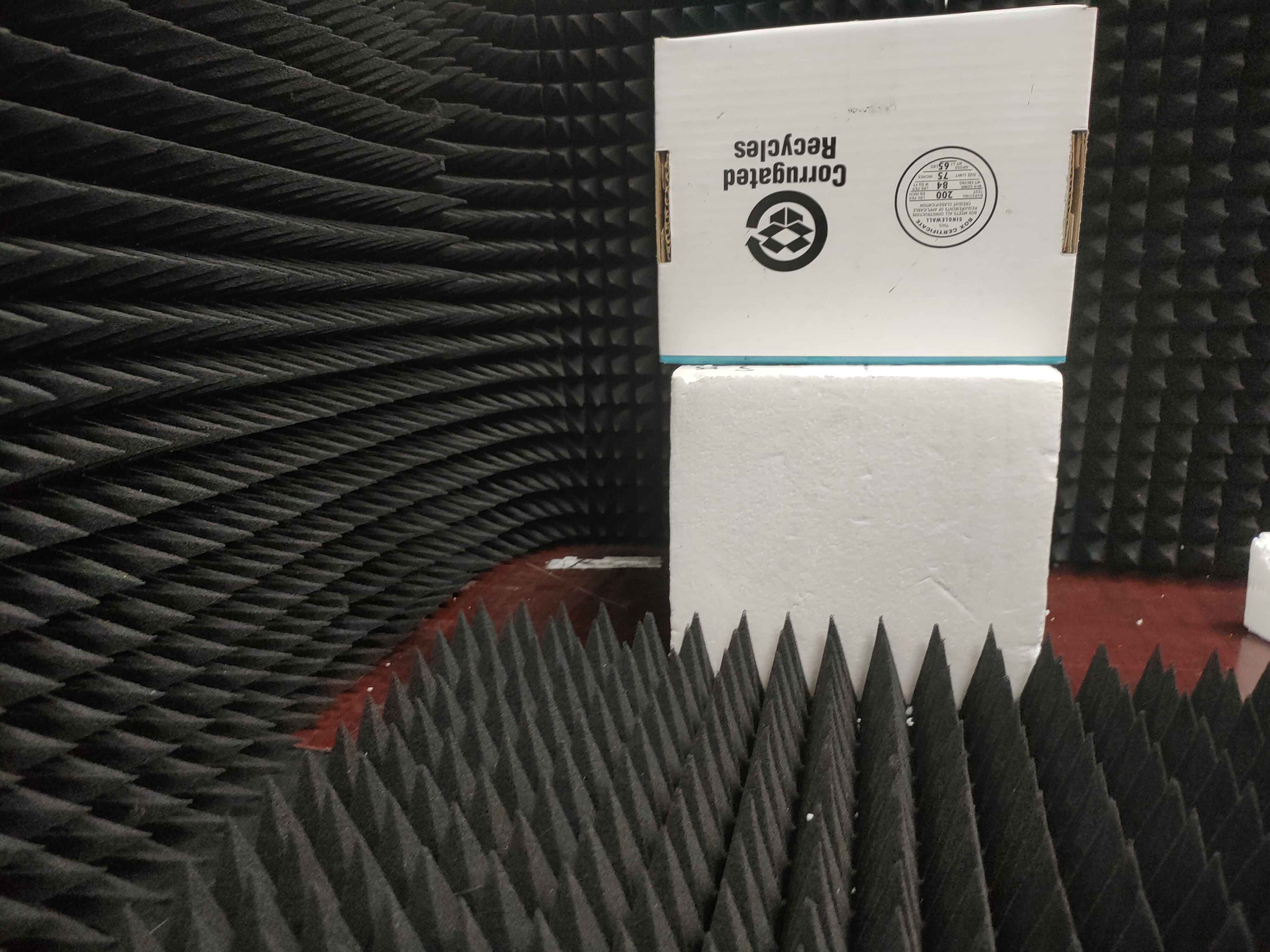}%
		\label{fig7_c}}
	\caption{The experimental scenario. (a) The aviation snip. (b) The hammer. (c) The box with a pistol.}
	\label{fig7}
\end{figure*} 

\begin{figure*}[!t]
	\centering
	\subfloat[]{\includegraphics[width=1.3in]{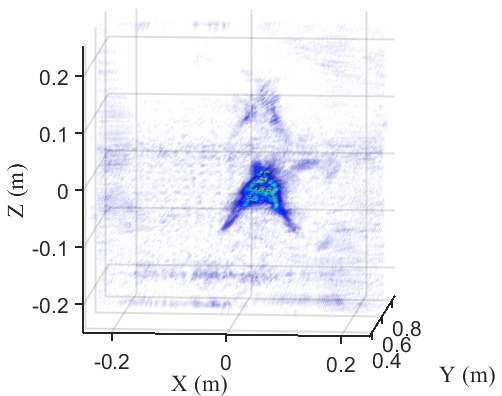}%
		\label{fig8_a}}
	\hfil
	\subfloat[]{\includegraphics[width=1.3in]{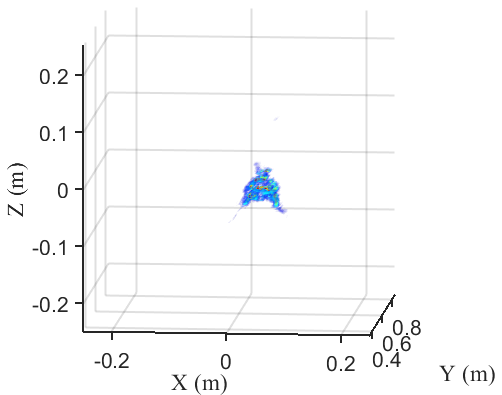}%
		\label{fig8_b}}
	\hfil
	\subfloat[]{\includegraphics[width=1.3in]{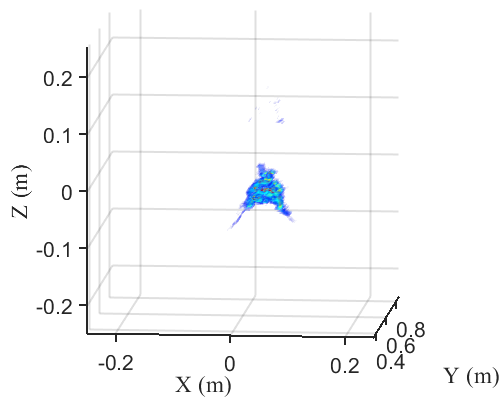}%
		\label{fig8_c}}
	\hfil
	\subfloat[]{\includegraphics[width=1.3in]{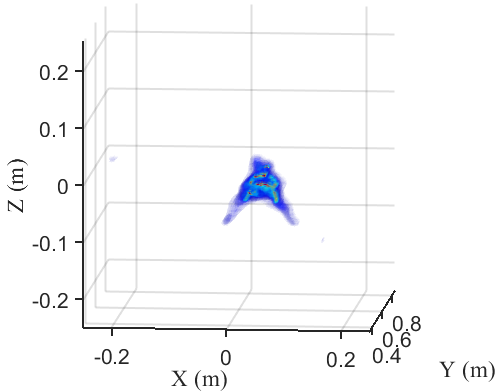}%
		\label{fig8_d}}
	\hfil
	\subfloat[]{\includegraphics[width=1.3in]{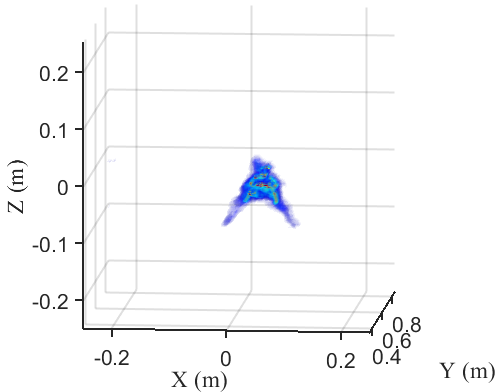}%
		\label{fig8_e}}
	\hfil
	\subfloat[]{\includegraphics[width=1.3in]{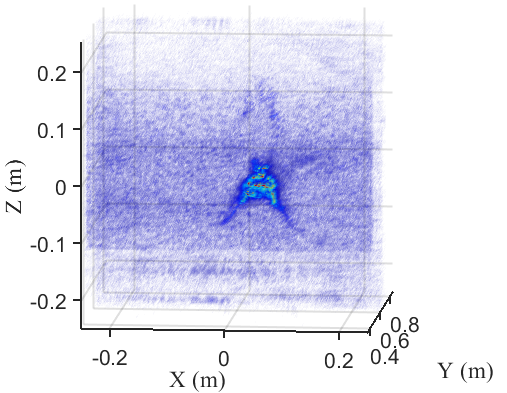}%
		\label{fig8_f}}
	\hfil
	\subfloat[]{\includegraphics[width=1.3in]{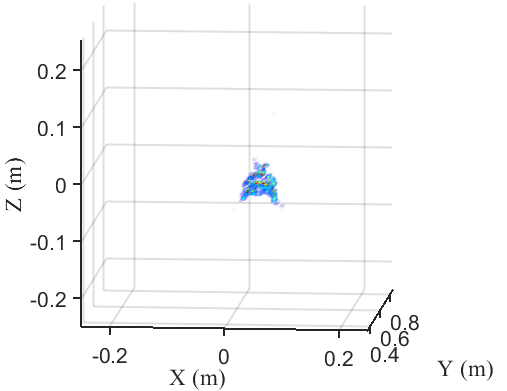}%
		\label{fig8_g}}
	\hfil
	\subfloat[]{\includegraphics[width=1.3in]{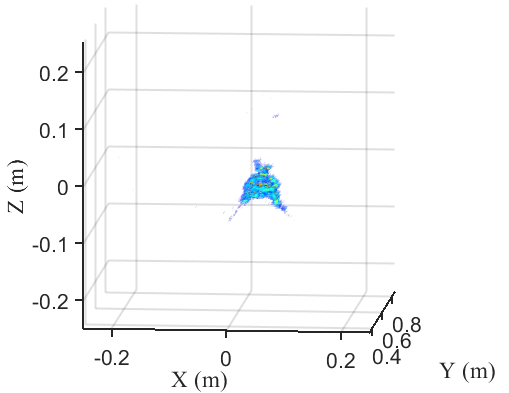}%
		\label{fig8_h}}
	\hfil
	\subfloat[]{\includegraphics[width=1.3in]{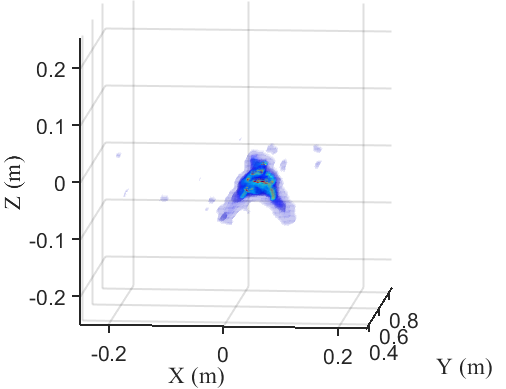}%
		\label{fig8_i}}
	\hfil
	\subfloat[]{\includegraphics[width=1.3in]{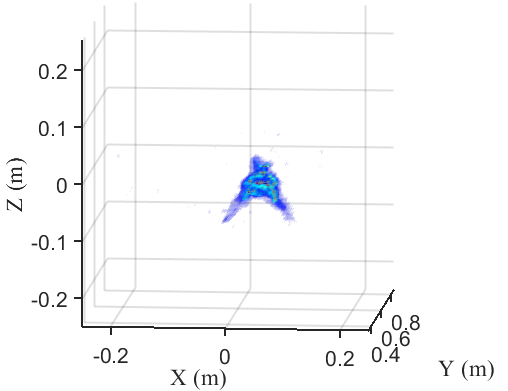}%
		\label{fig8_j}}
	\hfil
	\subfloat[]{\includegraphics[width=1.3in]{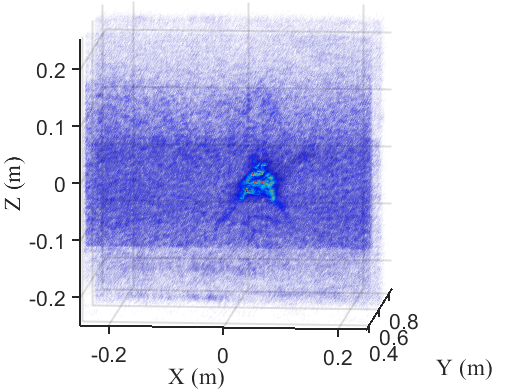}%
		\label{fig8_k}}
	\hfil
	\subfloat[]{\includegraphics[width=1.3in]{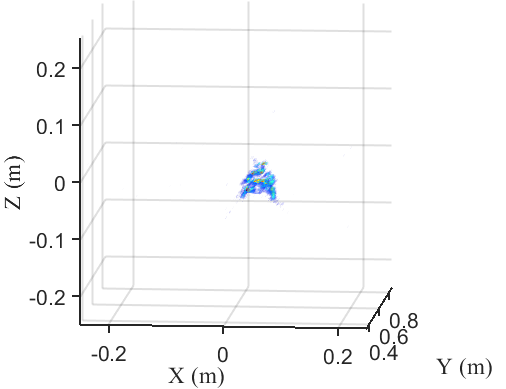}%
		\label{fig8_l}}
	\hfil
	\subfloat[]{\includegraphics[width=1.3in]{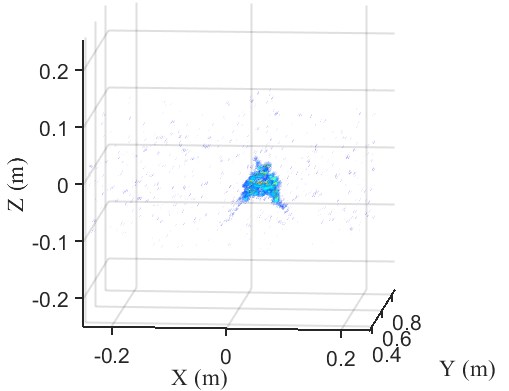}%
		\label{fig8_m}}
	\hfil
	\subfloat[]{\includegraphics[width=1.3in]{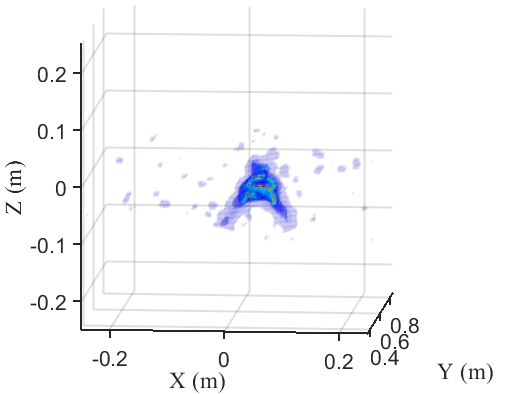}%
		\label{fig8_n}}
	\hfil
	\subfloat[]{\includegraphics[width=1.3in]{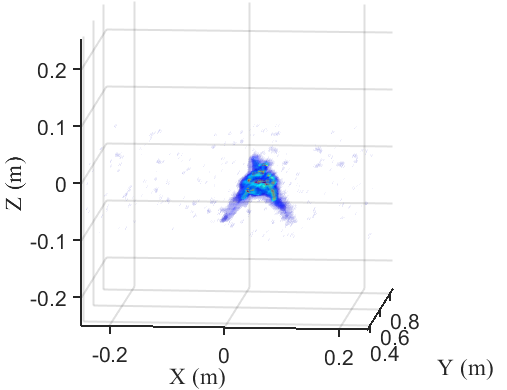}%
		\label{fig8_o}}
	\hfil
	\subfloat[]{\includegraphics[width=1.3in]{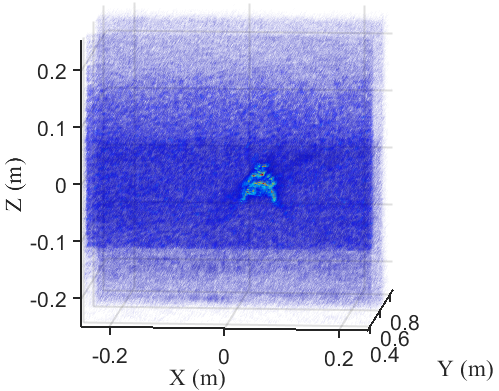}%
		\label{fig8_p}}
	\hfil
	\subfloat[]{\includegraphics[width=1.3in]{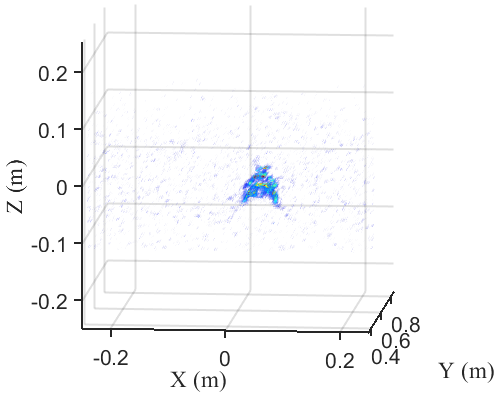}%
		\label{fig8_q}}
	\hfil
	\subfloat[]{\includegraphics[width=1.3in]{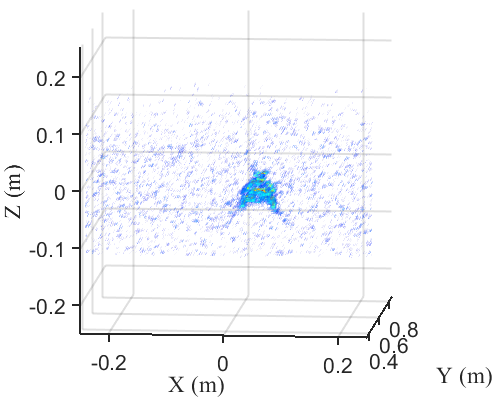}%
		\label{fig8_r}}
	\hfil
	\subfloat[]{\includegraphics[width=1.3in]{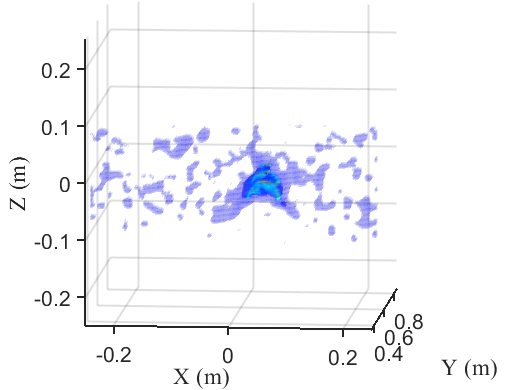}%
		\label{fig8_s}}
	\hfil
	\subfloat[]{\includegraphics[width=1.3in]{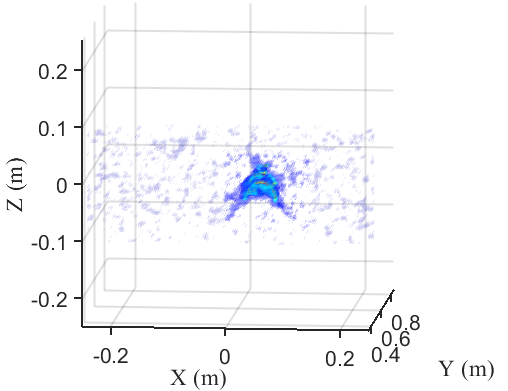}%
		\label{fig8_t}}
	\caption{The imaging results of the aviation snip at different SR. (a)-(e) The result of MF, $L_1$, GMCP, PnP and the proposed method at 75$\%$ SR. {(f)-(j) The result of MF, $L_1$, GMCP, PnP and the proposed method at 50$\%$ SR. (k)-(o) The result of MF, $L_1$, GMCP, PnP and the proposed method at 25$\%$ SR. (p)-(t) The result of MF, $L_1$, GMCP, PnP and the proposed method at 15$\%$ SR.}}
	\label{fig8}
\end{figure*} 

\begin{figure*}[!t]
	\centering
	\subfloat[]{\includegraphics[width=1.3in]{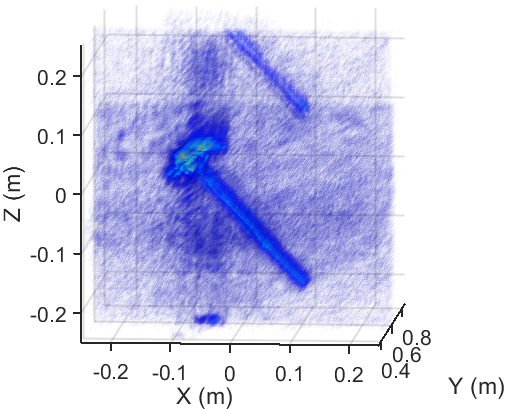}%
		\label{fig9_a}}
	\hfil
	\subfloat[]{\includegraphics[width=1.3in]{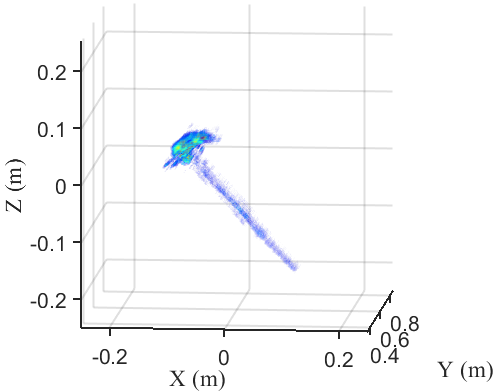}%
		\label{fig9_b}}
	\hfil
	\subfloat[]{\includegraphics[width=1.3in]{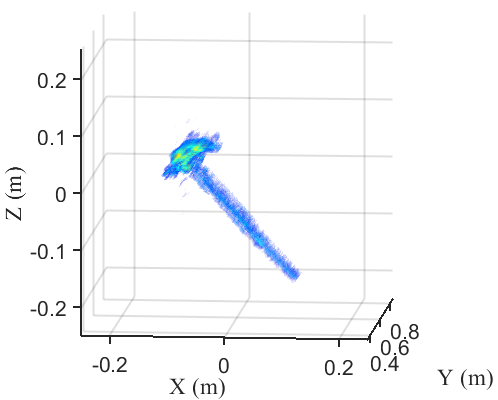}%
		\label{fig9_c}}
	\hfil
	\subfloat[]{\includegraphics[width=1.3in]{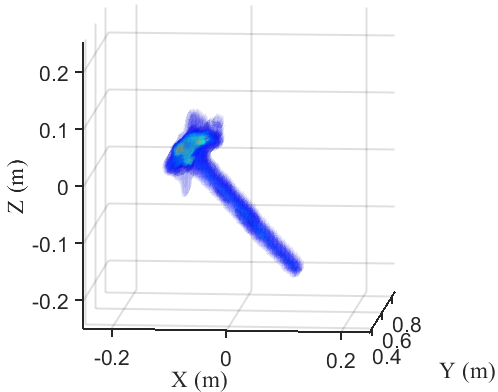}%
		\label{fig9_d}}
	\hfil
	\subfloat[]{\includegraphics[width=1.3in]{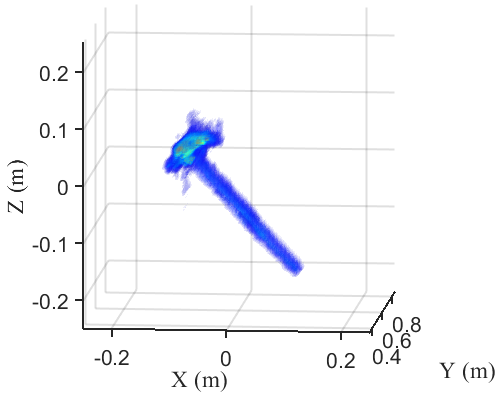}%
		\label{fig9_e}}
	\hfil
	\subfloat[]{\includegraphics[width=1.3in]{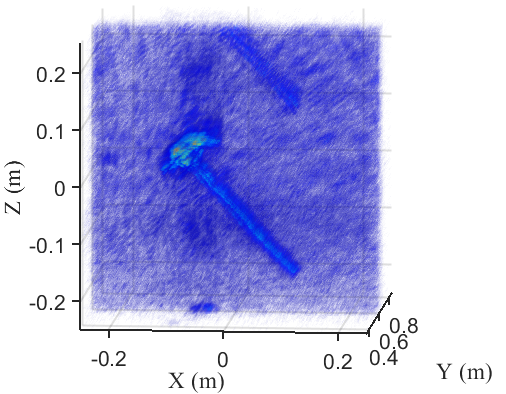}%
		\label{fig9_f}}
	\hfil
	\subfloat[]{\includegraphics[width=1.3in]{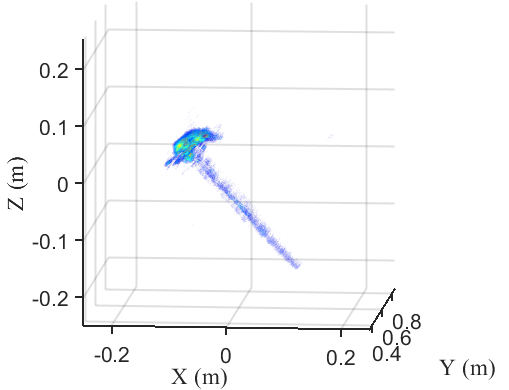}%
		\label{fig9_g}}
	\hfil
	\subfloat[]{\includegraphics[width=1.3in]{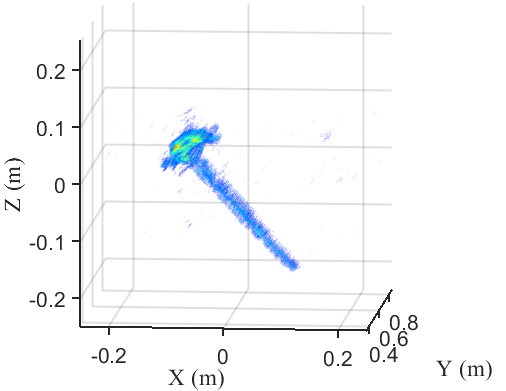}%
		\label{fig9_h}}
	\hfil
	\subfloat[]{\includegraphics[width=1.3in]{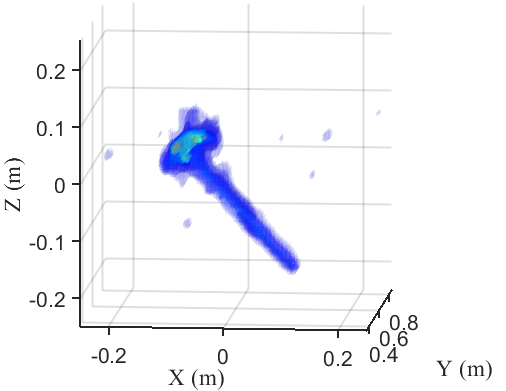}%
		\label{fig9_i}}
	\hfil
	\subfloat[]{\includegraphics[width=1.3in]{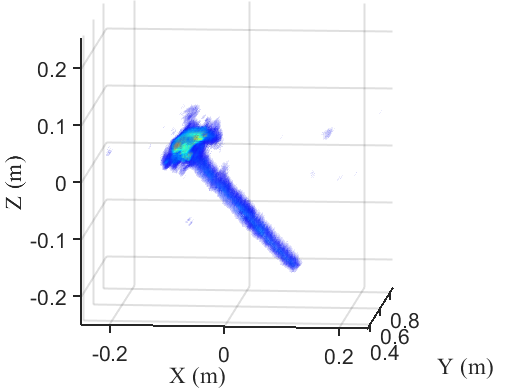}%
		\label{fig9_j}}
	\hfil
	\subfloat[]{\includegraphics[width=1.3in]{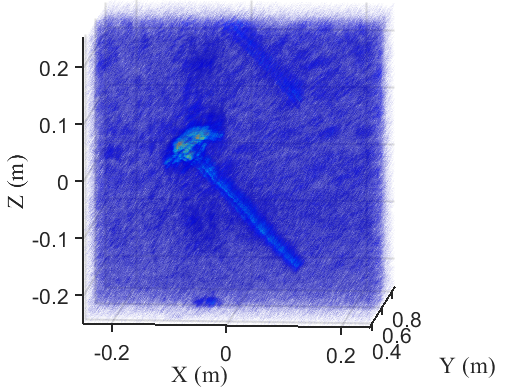}%
		\label{fig9_k}}
	\hfil
	\subfloat[]{\includegraphics[width=1.3in]{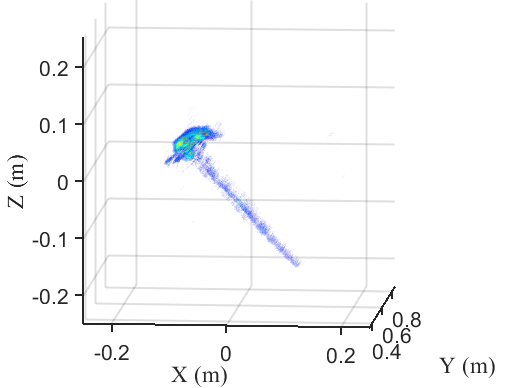}%
		\label{fig9_l}}
	\hfil
	\subfloat[]{\includegraphics[width=1.3in]{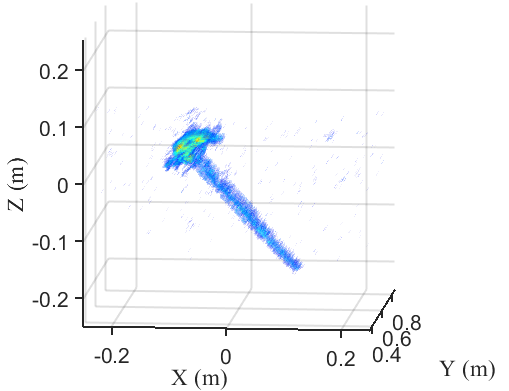}%
		\label{fig9_m}}
	\hfil
	\subfloat[]{\includegraphics[width=1.3in]{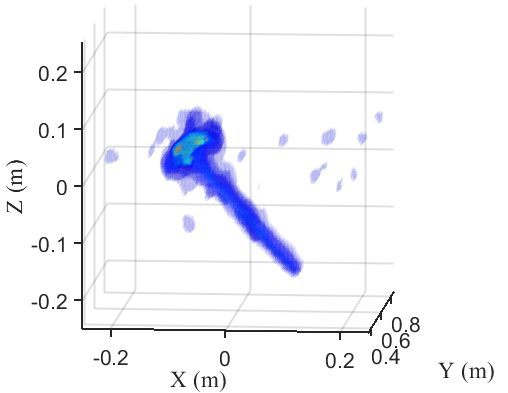}%
		\label{fig9_n}}
	\hfil
	\subfloat[]{\includegraphics[width=1.3in]{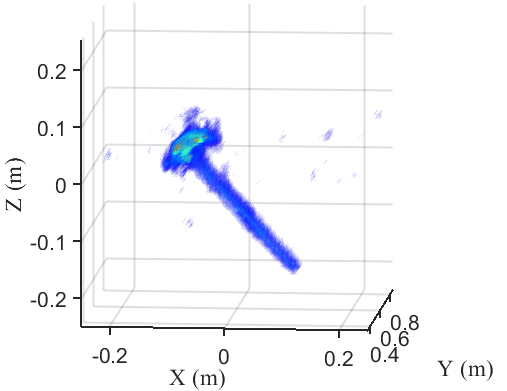}%
		\label{fig9_o}}
	\hfil
	\subfloat[]{\includegraphics[width=1.3in]{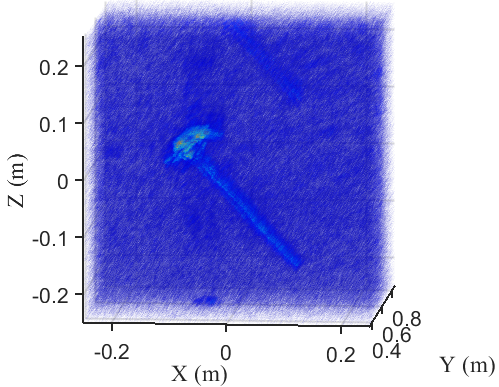}%
		\label{fig9_p}}
	\hfil
	\subfloat[]{\includegraphics[width=1.3in]{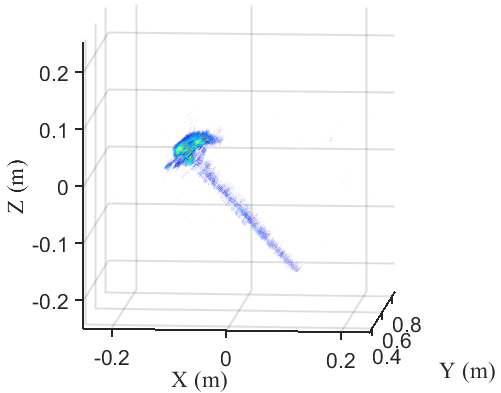}%
		\label{fig9_q}}
	\hfil
	\subfloat[]{\includegraphics[width=1.3in]{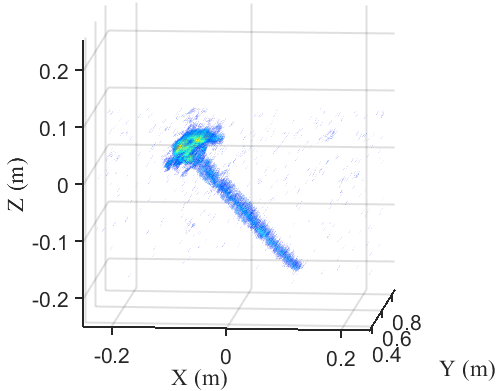}%
		\label{fig9_r}}
	\hfil
	\subfloat[]{\includegraphics[width=1.3in]{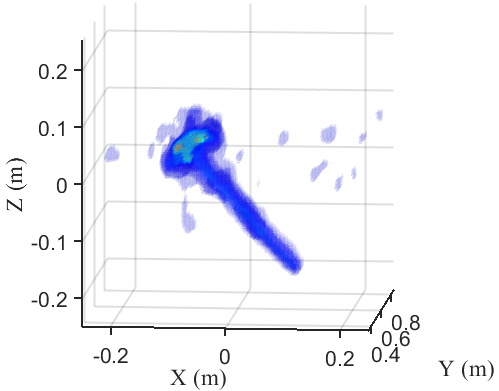}%
		\label{fig9_s}}
	\hfil
	\subfloat[]{\includegraphics[width=1.3in]{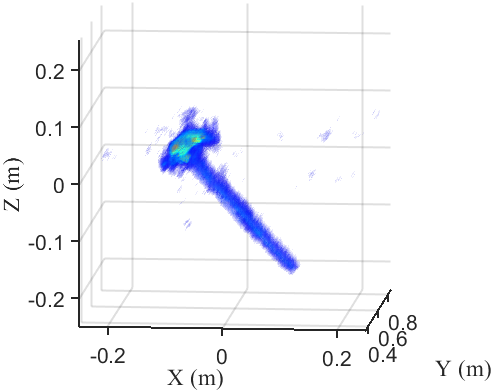}%
		\label{fig9_t}}
	\caption{The imaging results of the hammer at different SR. (a)-(e) The result of MF, $L_1$, GMCP, PnP and the proposed method at 75$\%$ SR. {(f)-(j) The result of MF, $L_1$, GMCP, PnP and the proposed method at 50$\%$ SR. (k)-(o) The result of MF, $L_1$, GMCP, PnP and the proposed method at 25$\%$ SR. (p)-(t) The result of MF, $L_1$, GMCP, PnP and the proposed method at 15$\%$ SR.}}
	\label{fig9}
\end{figure*}  

\begin{figure*}[!t]
	\centering
	\subfloat[]{\includegraphics[width=1.3in]{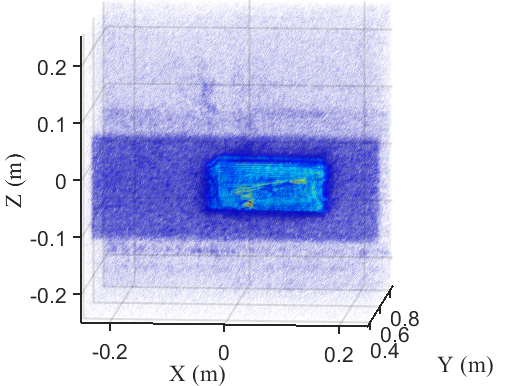}%
		\label{fig10_a}}
	\hfil
	\subfloat[]{\includegraphics[width=1.3in]{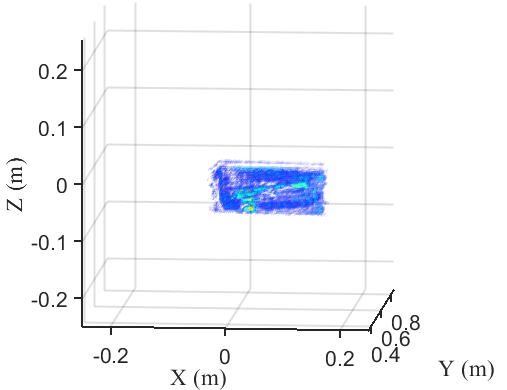}%
		\label{fig10_b}}
	\hfil
	\subfloat[]{\includegraphics[width=1.3in]{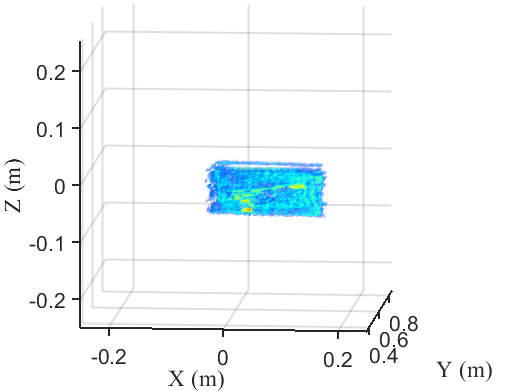}%
		\label{fig10_c}}
	\hfil
	\subfloat[]{\includegraphics[width=1.3in]{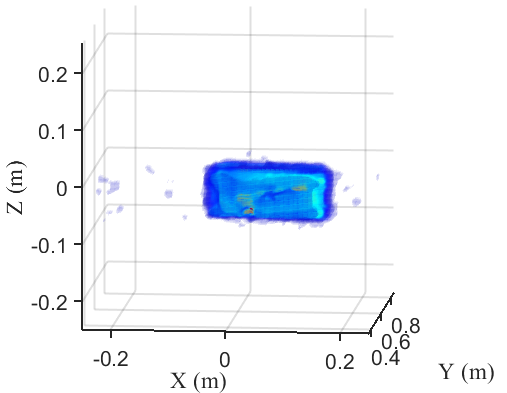}%
		\label{fig10_d}}
	\hfil
	\subfloat[]{\includegraphics[width=1.3in]{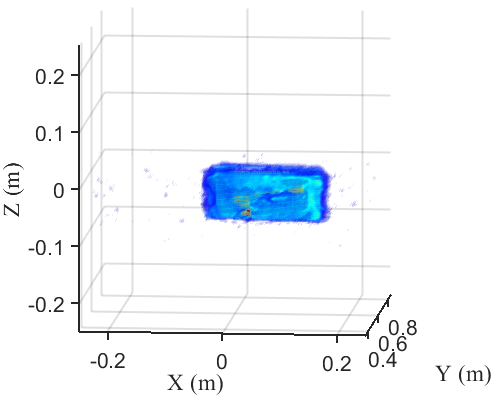}%
		\label{fig10_e}}
	\hfil
	\subfloat[]{\includegraphics[width=1.3in]{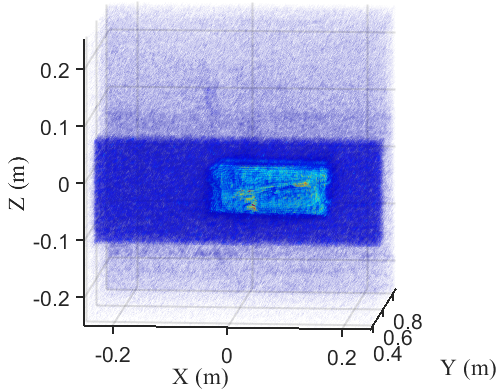}%
		\label{fig10_f}}
	\hfil
	\subfloat[]{\includegraphics[width=1.3in]{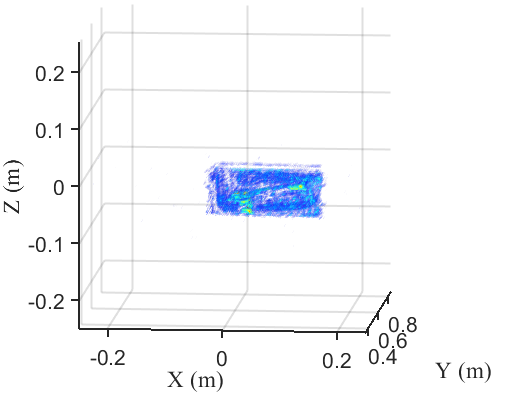}%
		\label{fig10_g}}
	\hfil
	\subfloat[]{\includegraphics[width=1.3in]{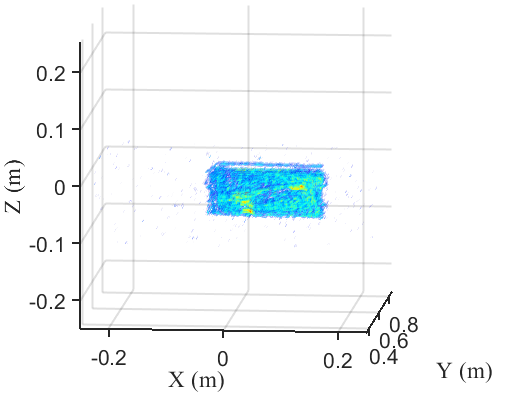}%
		\label{fig10_h}}
	\hfil
	\subfloat[]{\includegraphics[width=1.3in]{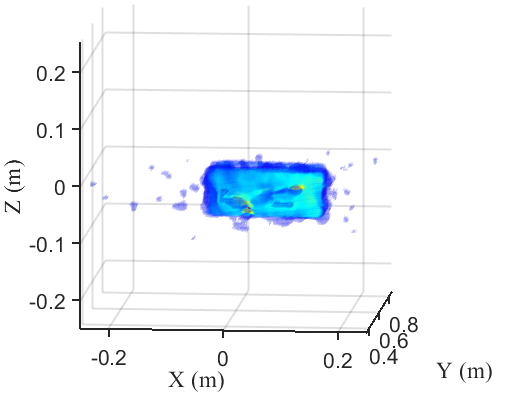}%
		\label{fig10_i}}
	\hfil
	\subfloat[]{\includegraphics[width=1.3in]{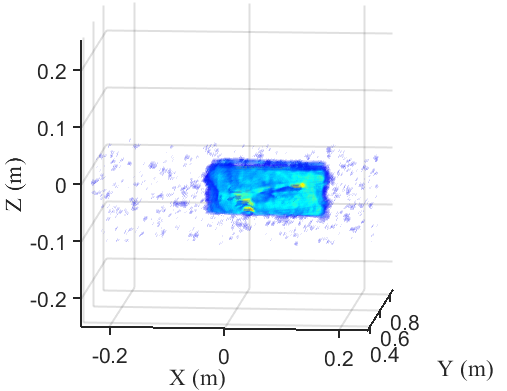}%
		\label{fig10_j}}
	\hfil
	\subfloat[]{\includegraphics[width=1.3in]{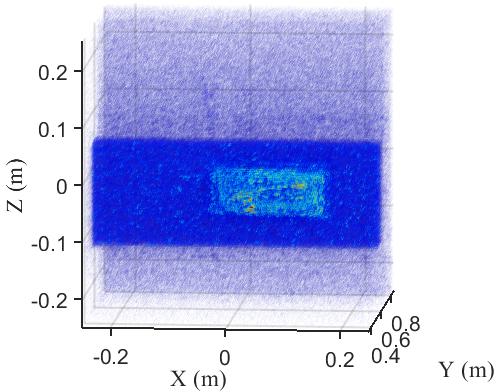}%
		\label{fig10_k}}
	\hfil
	\subfloat[]{\includegraphics[width=1.3in]{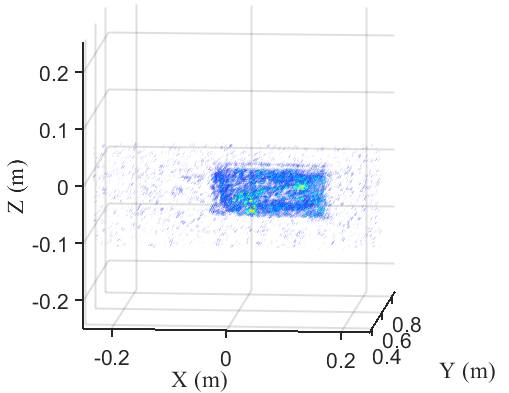}%
		\label{fig10_l}}
	\hfil
	\subfloat[]{\includegraphics[width=1.3in]{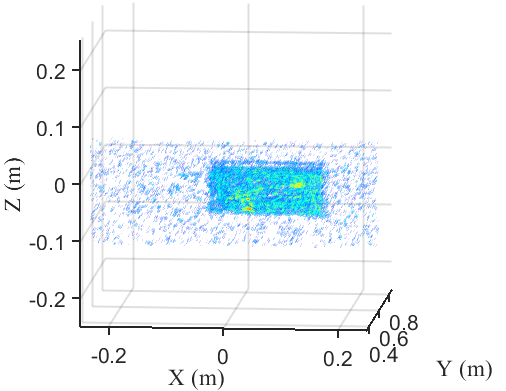}%
		\label{fig10_m}}
	\hfil
	\subfloat[]{\includegraphics[width=1.3in]{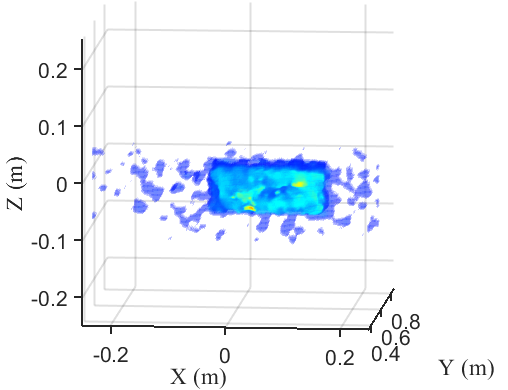}%
		\label{fig10_n}}
	\hfil
	\subfloat[]{\includegraphics[width=1.3in]{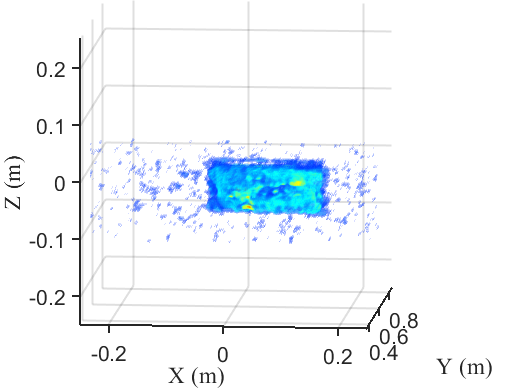}%
		\label{fig10_o}}
	\hfil
	\subfloat[]{\includegraphics[width=1.3in]{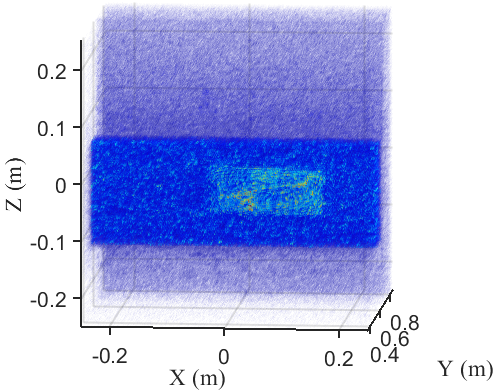}%
		\label{fig10_p}}
	\hfil
	\subfloat[]{\includegraphics[width=1.3in]{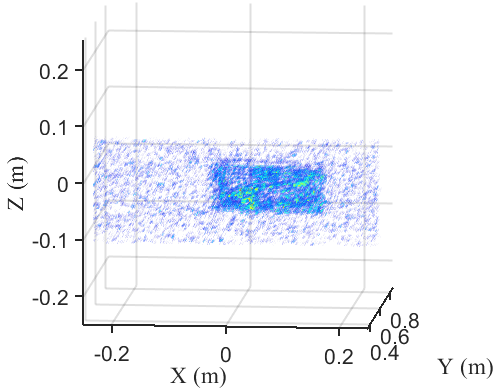}%
		\label{fig10_q}}
	\hfil
	\subfloat[]{\includegraphics[width=1.3in]{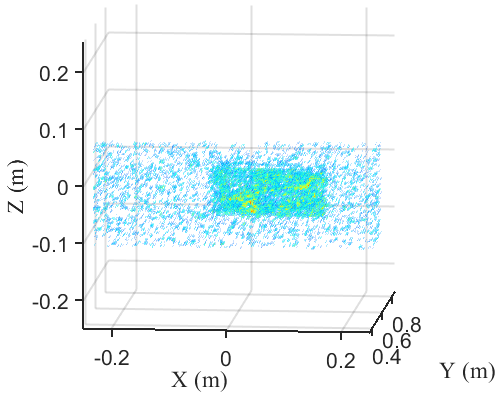}%
		\label{fig10_r}}
	\hfil
	\subfloat[]{\includegraphics[width=1.3in]{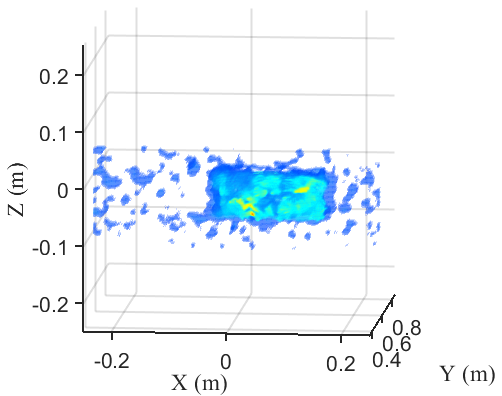}%
		\label{fig10_s}}
	\hfil
	\subfloat[]{\includegraphics[width=1.3in]{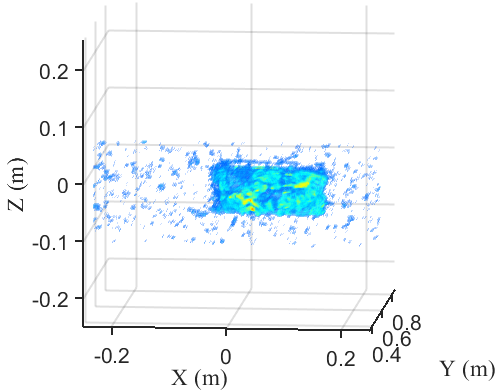}%
		\label{fig10_t}}
	\caption{The imaging results of the box with a pistol. (a)-(e) The result of MF, $L_1$, GMCP, PnP and the proposed method at 75$\%$ SR. {(f)-(j) The result of MF, $L_1$, GMCP, PnP and the proposed method at 50$\%$ SR. (k)-(o) The result of MF, $L_1$, GMCP, PnP and the proposed method at 25$\%$ SR. (p)-(t) The result of MF, $L_1$, GMCP, PnP and the proposed method at 15$\%$ SR.}}
	\label{fig10}
\end{figure*}

\begin{figure}[!t]
	\centering
	\subfloat[]{\includegraphics[width=1.6in]{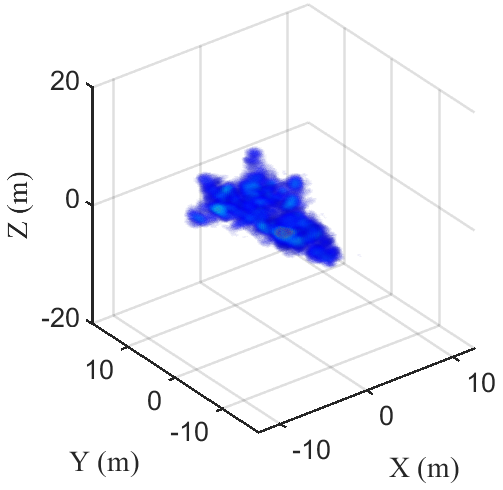}%
		\label{fig11_a}}
	\hfil
	\subfloat[]{\includegraphics[width=1.6in]{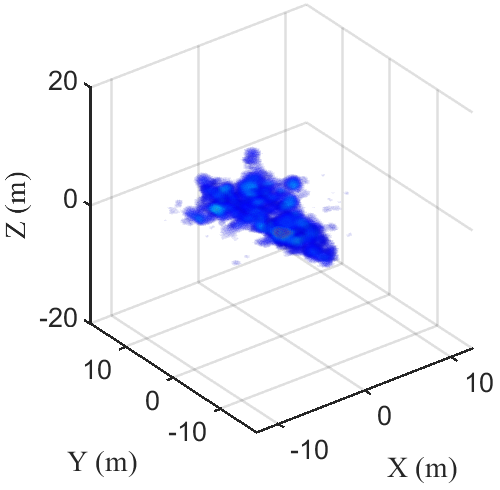}%
		\label{fig11_b}}
	\caption{{The imaging results of IRCNN. (a) The result at 50$\%$ sampling rate. (b) The result at 25$\%$ sampling rate.}}
	\label{fig11}
\end{figure}  

\subsection{Real Experiment of the Aviation Snip}

The measured SAR data of the aviation snip is used to validate the effectiveness of methods in real scenarios. Fig. 6(a) shows the experimental scenario of the aviation snip. The center frequency of the system is 77 GHz, signal bandwidth is 3.599 GHz, and array size is 0.4m × 0.4m. Fig. 7 shows the imaging results of the aviation snip under different SR. Compared with MF, sparse reconstruction methods can effectively improve the quality of 3D SAR images. 

The quantitative analysis of different methods is listed in Table~\ref{tab4}. At 75$\%$ SR, the PSNR of $L_1$, GMCP, PnP and the proposed method are 38.6498 dB, 40.9589 dB, 42.2122 dB and 44.0568 dB, respectively. The SSIM of $L_1$, GMCP, PnP and the proposed method are 0.7115, 0.7879, 0.8692 and 0.8838, respectively. {The NMSE of $L_1$, GMCP, PnP and the proposed method are 0.3909, 0.2848, 0.2638 and 0.1393, respectively. At 50$\%$ SR, the PSNR, SSIM and NMSE of the proposed method are 43.9651 dB, 0.8623 and 0.1812, respectively. At 25$\%$ SR, the PSNR, SSIM and NMSE of the proposed method are 40.8217 dB, 0.8398 and 0.2077, respectively. At 15$\%$ SR, the PSNR, SSIM and NMSE of the proposed method are 40.0013 dB, 0.9216 and 0.2259, respectively.} The results show that the proposed method can effectively improve SAR image quality and reconstruction accuracy in real scenes.

\begin{table*}[!t]
	\caption{PSNR, SSIM and NMSE of the aviation snip under different samplingbel\label{tab4}}
	\centering
	\begin{tabular}{|c|c|c|c|c|c|c|c|c|c|c|c|c|}
		\hline
		\multirow{2}{*}{SR} & \multicolumn{3}{c|}{$L_1$} & \multicolumn{3}{c|}{GMCP} & \multicolumn{3}{c|}{PnP} & \multicolumn{3}{c|}{RADMM}\\  
		\cline{2-13}
		& PSNR & SSIM & NMSE & PSNR & SSIM & NMSE & PSNR & SSIM & NMSE & PSNR & SSIM & NMSE\\
		\hline
		75 & 38.6498 & 0.7115 & 0.3909 & 40.9589 & 0.7879 & 0.2848 & 42.2122 & 0.8692 & 0.2638 & 44.0568 &  0.8838 & 0.1393\\
		\hline
		{50} & {35.6176} & {0.5478} & {0.6674} & {37.7830} & {0.5810}  & {0.5668} & {42.2014} & {0.8166} & {0.4415} & {43.9651} & {0.8623} & {0.1812} \\
		\hline
		{25} & {33.3788} & {0.4037} & {0.8213} & {35.4035} & {0.4354} & {0.7347} & {39.0650} & {0.7875} & {0.4879} & {40.8217} & {0.8398} & {0.2077}\\
		\hline
		{15} & {31.1881} & {0.2704} & {0.8738} & {33.4303} & {0.3255} & {0.7441} & {38.6257} & {0.8539} & {0.5120} & {40.0013} & {0.9216} & {0.2259}\\
		\hline
	\end{tabular}
\end{table*}

\subsection{Real Experiment of the Hammer}

The SAR data of the hammer is used to further analyze and verify the performance of the proposed method. Fig. 6(b) shows the experimental scenario. The main parameters of the system are as follows. The center frequency, signal bandwidth, and array size are 77 GHz, 3.599 GHz, and 0.4m × 0.4m, respectively. The 3D imaging result of the hammer at different SR is given in Fig. 8. The visualization result shows that the imaging result of MF contains significant noise and sidelobes, which are significantly suppressed through sparse reconstruction. Furthermore, the proposed method better preserves the feature information of the target, such as the information of the hammer handle. The quantitative analysis of different methods at different SR with PSNR, SSIM and NMSE is listed in Table~\ref{tab5}. {For 75$\%$ SR, the PSNR of $L_1$, GMCP, PnP and the proposed method are 31.1075 dB, 33.8051 dB, 38.7778 dB and 40.3129 dB, respectively. The SSIM of $L_1$, GMCP, PnP and the proposed method are 0.2779, 0.3702, 0.9253 and 0.9433, respectively. The NMSE of $L_1$, GMCP, PnP and the proposed method are 0.6034, 0.4533, 0.2056 and 0.1527, respectively. The PSNR and SSIM of the proposed method are the highest among these methods. The NMSE of RADMM is the lowest. At low SR, the PSNR and SSIM of the proposed method are still higher than MF, $L_1$, GMCP and PnP, validating that the proposed method has high robustness.}

\begin{table*}[!t]
	\caption{PSNR and SSIM of the hammer under different samplingbel\label{tab5}}
	\centering
	\begin{tabular}{|c|c|c|c|c|c|c|c|c|c|c|c|c|}
		\hline
		\multirow{2}{*}{SR} & \multicolumn{3}{c|}{$L_1$} & \multicolumn{3}{c|}{GMCP} & \multicolumn{3}{c|}{PnP} & \multicolumn{3}{c|}{RADMM}\\  
		\cline{2-13}
		& PSNR & SSIM & NMSE & PSNR & SSIM & NMSE & PSNR & SSIM & NMSE & PSNR & SSIM & NMSE\\
		\hline
		75 & 31.1075 & 0.2779 & 0.6034 & 33.8051 & 0.3702 & 0.4533 & 38.7778 & 0.9253 & 0.2056 & 40.3129 &  0.9433 & 0.1527\\
		\hline
		{50} & {29.0606} & {0.1705} & {0.6994} & {31.3226} & {0.2514} & {0.5805} & {36.1537} & {0.8585} & {0.3525} & {39.1204} & {0.9238} & {0.1748}\\
		\hline
		{25} & {28.0885} & {0.1299} & {0.7507} & {29.9337} & {0.2015} & {0.6381} & {34.5722} & {0.8047} & {0.4318} & {37.6809} & {0.8968} & {0.2542}\\
		\hline
		{15} & {27.1908} & {0.0969} & {0.7967} & {29.0691} & {0.1596} & {0.6974} & {32.9824} & {0.7440} & {0.4901} & {34.0490} & {0.8108} & {0.3166}\\
		\hline
	\end{tabular}
\end{table*}

\subsection{Real Experiment of the Box with a Pistol}

We perform imaging processing on real SAR data of the box with a pistol to verify the effectiveness of the proposed method in complex situations. {The center frequency, signal bandwidth, and array size are 77 GHz, 3.599 GHz, and 0.4m × 0.4m, respectively.} Fig. 6(c) shows the experimental scenario of the box with a pistol. {Fig. 9 (a)-(e) shows the 3D imaging result at 75$\%$ SR. Fig. 9 (f)-(j) shows the 3D imaging result at 50$\%$ SR. Fig. 9 (k)-(o) shows the 3D imaging result at 25$\%$ SR. Fig. 9 (p)-(t) shows the 3D imaging result at 15$\%$ SR.} It can be seen that the proposed method effectively suppressing sidelobes while preserves the contour information of the target. Quantitative analysis is listed in Table~\ref{tab6}. {For 75$\%$ SR, the PSNR of $L_1$, GMCP, PnP and the proposed method are 29.4996 dB, 33.8427 dB, 37.8787 dB and 38.8004 dB, respectively. The SSIM of $L_1$, GMCP, PnP and the proposed method are 0.2784, 0.4490, 0.8936 and 0.9083, respectively. The NMSE of $L_1$, GMCP, PnP and the proposed method are 0.6403, 0.3685, 0.2222 and 0.1821, respectively.} At low sampling, the PSNR and SSIM of the proposed method are also the highest among these methods and the NMSE of the proposed method is lowest. The results of different SAR experiments show that the proposed method effectively improve the quality of 3D SAR images and preserve contour information in real environments. 

\begin{table*}[!t]
	\caption{PSNR and SSIM of the box with a pistol under different samplingbel\label{tab6}}
	\centering
	\begin{tabular}{|c|c|c|c|c|c|c|c|c|c|c|c|c|}
		\hline
		\multirow{2}{*}{SR} & \multicolumn{3}{c|}{$L_1$} & \multicolumn{3}{c|}{GMCP} & \multicolumn{3}{c|}{PnP} & \multicolumn{3}{c|}{RADMM}\\  
		\cline{2-13}
		& PSNR & SSIM & NMSE & PSNR & SSIM & NMSE & PSNR & SSIM & NMSE & PSNR & SSIM & NMSE\\
		\hline
		75 & 29.4996 & 0.2784 & 0.6403 & 33.8427 & 0.4490 & 0.3685 & 37.8787 & 0.8936 & 0.2222 & 38.8004 &  0.9083 & 0.1821\\
		\hline
		{50} & {28.2343} & {0.2138} & {0.6919} & {32.1447} & {0.3655} & {0.4554} & {36.0330} & {0.8754} & {0.3754} & {38.3559} & {0.8995} & {0.1961}\\
		\hline
		{25} & {26.6048} & {0.1486} & {0.7543} & {30.2657} & {0.2913} & {0.5352} & {33.6086} & {0.7617} & {0.4574} & {36.2380} & {0.8613} & {0.2058}\\
		\hline
		{15} & {25.5755} & {0.1250} & {0.7894} & {28.8778} & {0.2645} & {0.5868} & {32.7171} & {0.7495} & {0.4708} & {34.2471} & {0.8375} & {0.2212}\\
		\hline
	\end{tabular}
\end{table*} 

\section{Discussion}
{In this article, the proposed framework utilizes denoisers to construct explicit priors, improving the accuracy of SAR sparse imaging. Compared with traditional regularization functions such as $L_1$ and GMCP, the proposed method can better reconstruct target information even under few observed SAR data. Compared with PnP, the proposed method has superior sparse imaging performance at low SR or low SNR.

To further verify the flexibility and generality of the proposed framework, the deep denoiser IRCNN \cite{refz6} is initially combined to achieve SAR 3D sparse imaging. The 3D imaging results of IRCNN are shown in Fig. 10, which indicates that even at lower SR, the proposed framework can integrate deep denoiser to achieve superior sparse reconstruction and effectively preserves target information. For 50$\%$ SR, PSNR, SSIM and NMSE are 48.5432 dB, 0.9873 and 0.0778, respectively. For 25$\%$ SR, PSNR, SSIM and NMSE are 48.1014 dB, 0.9757 and 0.1113, respectively. Quantitative analysis of other methods is listed in Table~\ref{tab1}. The results indicate that the proposed method can achieve superior sparse SAR imaging performance. Additionally, the proposed framework can also be integrated with existing regularization functions to achieve sparse imaging, which also demonstrates the generality of the proposed framework. For computational complexity, the proposed framework can integrate linearization ideas and imaging kernel functions \cite{ref12} to further reduce the required computational complexity.
}

\section{Conclusion}
{A flexible sparse imaging framework based on RED and proximal gradient descent type method is proposed to achieve high-precision 3D SAR images in this article.} Firstly, to address the issue of insufficient target feature representation of handcrafted regularization functions in 3D SAR imaging, we utilize state-of-the-art denoising operators to construct explicit prior terms. {The proposed method better enhance the features of the target and preserve the structure information of the target compared to existing 3D SAR sparse imaging method under few observed SAR data. Then, different iterative algorithms are integrated to verify the generality of the proposed framework, such as ADMM and GAP. The proposed framework can be applied to high-dimensional imaging scenes and can be integrated with different regularization functions and denoisers to achieve sparse imaging, which further verify the flexibility and generality of the proposed framework.} Additionally, the proposed framework has an explicit objective function and robust convergence, making it suitable for {for few observed SAR data, such as low SR or SNR}. Extensive experiments based on simulation and real SAR data are implemented to verify the effectiveness and generality of the proposed method. Quantitative and visual analysis show that the proposed method has better sparse reconstruction performance compared to the regularization functions studied and PnP.


\newpage

\begin{IEEEbiographynophoto}{Yangyang Wang}
pursued the M.S. degree from University of Electronic Science and Technology of China (UESTC), Chengdu, China, from 2016 to 2018, and received the Ph.D. degrees from UESTC in 2022. In 2023, he joined North University of China (NUC).

His current research interests include three-dimensional synthetic aperture radar (SAR) imaging, sparse signal processing and radar cross section (RCS) measurement.
\end{IEEEbiographynophoto}

\begin{IEEEbiographynophoto}{Xu Zhan}
	received the B.Sc. degree from School of Information and Communication Engineering, University of Electronic Science and Technology of China (UESTC), Chengdu, China in 2018, where he is currently pursing the Ph.D. degree.
	
	His research focus includes sparse-reconstruction methods and their application in 3D linear-array synthetic aperture radar (LA-SAR), radar cross section (RCS) measurement, interferometric SAR (InSAR) or multibaseline SAR (TomoSAR), azimuth multichannel SAR (AMC-SAR), velocity SAR (VSAR) and video SAR (ViSAR).
\end{IEEEbiographynophoto}

\begin{IEEEbiographynophoto}{Jing Gao}  is currently pursuing the M.S. degree with the School of Information and Communication Engineering, North University of China (NUC), Taiyuan, China.
	
	Her research interests include sparse synthetic aperture radar (SAR) imaging, radar target detection and signal processing.
\end{IEEEbiographynophoto}

\begin{IEEEbiographynophoto}{Jinjie Yao}
	received the Ph.D. degree in North University of China (NUC), China, in 2011. Since 2011, he was an Associate Professor with NUC. His research interests include Microwave and Millimeter Wave Technology and Electromagnetic Detection Technology.
\end{IEEEbiographynophoto}

\begin{IEEEbiographynophoto}{Shunjun Wei}
	received the B.S., M.Sc., and Ph.D. degrees in electronic engineering from the University of Electronic Science and Technology of China (UESTC), Chengdu, China,
	in 2006, 2009, and 2013, respectively.
	
	In 2014, he joined UESTC, where he is currently an Professor. His research interests include radar signal processing, spares imaging and SAR systems.
\end{IEEEbiographynophoto}

\begin{IEEEbiographynophoto}{Jiansheng Bai}
	received the M.S.degree from North University of China (NUC), Taiyuan, China, where he is currently pursuing the Ph.D. degree with the School of Information and Communication Engineering. His research interests include laser detection, EM situational awareness and signal processing.
\end{IEEEbiographynophoto}

\vfill


\begin{thebibliography}{}
\providecommand{\url}[1]{#1}
\csname url@samestyle\endcsname
\providecommand{\newblock}{\relax}
\providecommand{\bibinfo}[2]{#2}
\providecommand{\BIBentrySTDinterwordspacing}{\spaceskip=0pt\relax}
\providecommand{\BIBentryALTinterwordstretchfactor}{4}
\providecommand{\BIBentryALTinterwordspacing}{\spaceskip=\fontdimen2\font plus
\BIBentryALTinterwordstretchfactor\fontdimen3\font minus \fontdimen4\font\relax}
\providecommand{\BIBforeignlanguage}[2]{{%
\expandafter\ifx\csname l@#1\endcsname\relax
\typeout{** WARNING: IEEEtran.bst: No hyphenation pattern has been}%
\typeout{** loaded for the language `#1'. Using the pattern for}%
\typeout{** the default language instead.}%
\else
\language=\csname l@#1\endcsname
\fi
#2}}
\providecommand{\BIBdecl}{\relax}
\BIBdecl

\end{thebibliography}


\begin{thebibliography}{1}
\bibliographystyle{IEEEtran}

\bibitem{ref1}
M. Cetin, Mujdat, and W. C. Karl, ``Feature-enhanced synthetic aperture radar image formation based on nonquadratic regularization,'' \textit{IEEE Trans. Image Process.}, vol. 10, no. 4, pp. 623--631, 2001.

\bibitem{ref2}
A. Moreira, P. Prats-Iraola, M. Younis, G. Krieger, I. Hajnsek, and K. P. Papathanassiou, ``A tutorial on synthetic aperture radar,'' \textit{IEEE Geosci. Remote Sens. Mag.}, vol. 1, no. 1, pp. 6--43, 2013.

\bibitem{ref3}
Z. Xu, M. Liu, G. Zhou, Z. Wei, B. Zhang and Y. Wu, ``An accurate sparse SAR imaging method for enhancing region-based features via nonconvex and TV regularization,'' \textit{IEEE J. Sel. Top. Appl. Earth Observ. Remote Sens.}, vol. 14, no. 3, pp. 235--250, 2021.

\bibitem{ref4}
Z. Wang, Q. Guo, X. Tian, T. Chang and H. -L. Cui, ``Near-Field 3-D millimeter-wave imaging using MIMO RMA with range compensation,'' \textit{IEEE Trans. Microw. Theory Tech.}, vol. 67, no. 3, pp. 1157--1166, 2019.

\bibitem{ref5}
S. Wei et al., ``3DRIED: A high-resolution 3-D millimeter-wave radar dataset dedicated to imaging and evaluation,'' \textit{Remote Sens.}, vol. 13, no. 7, pp. 3366, 2021.

\bibitem{ref6}
Y. Wang et al., ``An RCS measurement method using sparse imaging based 3-D SAR complex image,'' \textit{IEEE Antennas Wirel. Propag. Lett.}, vol. 21, no. 1, pp. 24–-28, 2022.

\bibitem{refs1}
T. Zeng et al., ``Unsupervised 3D array-SAR imaging based on generative model for scattering diagnosis,'' \textit{IEEE Antennas Wirel. Propag. Lett.}, 2024, doi: 10.1109/LAWP.2024.3395771.

\bibitem{ref7}
M. Wang, S. Wei, Z. Zhou, et al., ``3-D SAR autofocusing with learned sparsity,'' \textit{IEEE Trans. Geosci. Remote Sens.}, vol. 60, pp. 1--18, 2022.

\bibitem{ref8}
A. Shakeel, R. J. Walters, S. K. Ebmeier, and N. A. Moubayed, ``ALADDIn: Autoencoder-LSTM-based anomaly detector of deformation in InSAR,'' \textit{IEEE Trans. Geosci. Remote Sens.}, vol. 60, 2022, Art. no. 4706512.

\bibitem{ref9}
X. X. Zhu, S. Montazeri, C. Gisinger, R. F. Hanssen, and R. Bamler, ``Geodetic SAR tomography,'' \textit{IEEE Trans. Geosci. Remote Sens.}, vol. 54, no. 1, pp. 18–35, 2016.

\bibitem{ref10}
K. Liao, X. Zhang, and J. Shi, ``Plane-wave synthesis and RCS extraction via 3-D linear array SAR,'' \textit{IEEE Antennas Wireless Propag. Lett.}, vol. 14, pp. 994–-997, 2015.

\bibitem{ref11}
M. Wang, S. Wei, J. Liang; S. Liu, J. Shi, X. Zhang, ``Lightweight FISTA-inspired sparse reconstruction network for mmW 3-D holography,'' \textit{IEEE Trans. Geosci. Remote Sens.}, vol. 60, pp. 1–-20, 2022.

\bibitem{ref12}
Y. Wang, Z. He, X. Zhan, Y. Fu, and L. Zhou, ``Three-dimensional sparse SAR imaging with generalized $L_q$ regularization,'' \textit{Remote Sens.}, vol. 14, no. 2, pp. 288, 2022.

\bibitem{ref13}
L. M. H. Ulander, H. Hellsten, G. Stenstrom, ``Synthetic-aperture radar processing using fast factorized back-projection,'' \textit{IEEE Trans. Aerosp. Electron. Syst.}, vol. 39, no. 3, pp. 760-776, 2003.

\bibitem{ref14}
A. S. Milman, ``SAR imaging by wk migration,'' \textit{Int. J. Remote Sens.}, vol. 14, no. 10, pp. 1965-1979, 1993.

\bibitem{ref15}
O. Tulgar, and A A. Ergin, ``Improved pencil back-projection method with image segmentation for farfield/near-field SAR imaging and RCS extraction,'' \textit{IEEE Trans. Antennas Propag.}, vol. 63, no. 6, pp. 2572--2584, 2015.

\bibitem{ref16}
O. Karakus¸, P. Mayo and A. Achim, ``Convergence guarantees for nonconvex optimisation with Cauchy-based penalties,'' \textit{IEEE Trans. Signal Process.}, vol. 68, pp. 6159-–6170, 2020.

\bibitem{ref17}
H. Bi, B. Zhang, X. X. Zhu, J. Sun, W. Hong, and Y. Wu, ``$L_1$-regularization-based SAR imaging and CFAR detection via complex approximated message passing,'' \textit{IEEE Trans. Geosci. Remote Sens.}, vol. 55, no. 6, pp. 3426–-3440, 2017.

\bibitem{ref18}
S. Osher, F. Ruan, J. Xiong, Y. Yao, and W. Yin, ``Sparse recovery via differential inclusions,'' \textit{Appl. Comput. Harmon. Anal.}, vol. 41, no. 2, pp. 436–-469, 2016.

\bibitem{ref19}
M. Li, J. Wu, W. Huo, Z. Li, J. Yang and H. Li, ``STLS-LADMM-Net: A deep network for SAR autofocus imaging,'' \textit{IEEE Trans. Geosci. Remote Sens.}, vol. 60, pp. 1--14, 2022.

\bibitem{ref20}
Y. Wang, Z. He, F. Yang, Q. Zeng, and X. Zhan, ``3D sparse SAR image reconstruction based on Cauchy penalty and convex optimization,'' \textit{Remote Sens.}, vol. 14, no. 10, pp. 2308, 2022.

\bibitem{ref21}
A. Wu, Y. Wu, Y. Jin, Y. Wang and Z. Guo, ``$L_{1/2}$ regularization for ISAR imaging and target enhancement of complex image,'' \textit{IEEE Trans. Geosci. Remote Sens.}, vol. 60, pp. 1--10, 2022.

\bibitem{ref22}
G. Xu, B. Zhang, J. Chen, F. Wu, J. Sheng, and W. Hong, ``Sparse inverse synthetic aperture radar imaging using structured low-rank method,'' \textit{IEEE Trans. Geosci. Remote Sens.}, vol. 60, pp. 1–-12, 2022.

\bibitem{ref23}
J. Fan and R. Li, ``Variable selection via nonconcave penalized likelihood and its oracle properties,'' \textit{J. Am. Stat. Assoc.}, vol. 96, no. 456, pp. 1348-–1360, 2001.

\bibitem{ref24}
C.-H. Zhang et al., ``Nearly unbiased variable selection under minimax concave penalty,'' \textit{The Annals of statistics.}, vol. 38, no. 2, pp. 894–-942, 2010.

\bibitem{ref25}
O. Karakus¸ and A. Achim, ``On solving SAR imaging inverse problems using nonconvex regularization with a Cauchy-based penalty,'' \textit{IEEE Trans. Geosci. Remote Sens.}, vol. 59, n. 7, pp. 5828–-5840, 2021.

\bibitem{refz1}
{X. Liao, H. Li, and L. Carin, ``Generalized alternating projection for weighted-$l_{2,1}$ minimization with applications to model-based compressive sensing,'' \textit{SIAM J. Imag. Sci.}, vol. 7, n. 2, pp. 797–-823, 2014.}

\bibitem{ref26}
D. L. Donoho, A. Maleki, and A. Montanari, ``Message-passing algorithms for compressed sensing,'' \textit{Proc. Nat. Acad. Sci.}, vol. 106, no. 45, pp. 18914-–18919, 2009.

\bibitem{ref27}
I. Daubechies, M. Defriese; C. De Mol, ``An iterative thresholding algorithm for linear inverse problems with a sparsity constraint,'' \textit{Commun. Pure Appl. Math.}, vol. 57, pp. 1413–-1457, 2004.

\bibitem{ref28}
S. Boyd, et al., ``Distributed optimization and statistical learning via the alternating direction method of multipliers,'' \textit{Found. Trends Mach. Learn.}, vol. 13.1, pp. 1--122, 2011.

\bibitem{ref29}
M. V. Afonso, J. M. Bioucas-Dias and M. A. T. Figueiredo, ``An augmented lagrangian approach to the constrained optimization formulation of imaging inverse problems,'' \textit{IEEE Trans. Image Process.}, vol. 20, no. 3, pp. 681--695, 2011.

\bibitem{ref30}
A. Güngör, M. Çetin and H. E. Güven, ``Compressive synthetic aperture radar imaging and autofocusing by augmented Lagrangian methods,'' \textit{IEEE Trans. Comput. Imaging}, vol. 8, pp. 273--285, 2022.

\bibitem{ref31}
H. Bi, X. Lu, Y. Yin, W. Yang and D. Zhu, ``Sparse SAR imaging based on periodic block sampling data,'' \textit{IEEE Trans. Geosci. Remote Sens.}, vol. 60, pp. 1--12, 2022.

\bibitem{ref32}
S. V. Venkatakrishnan, C. A. Bouman, and B. Wohlberg, ``Plug-and-Play priors for model based reconstruction,'' in \textit{Proc. IEEE Global Conf. Signal Inf. Process.}, Austin, TX, USA, 2013, pp. 945–948.

\bibitem{ref33}
S. H. Chan, X. Wang and O. A. Elgendy, ``Plug-and-Play ADMM for image restoration: fixed-point convergence and applications,'' \textit{IEEE Trans. Comput. Imaging}, vol. 3, no. 1, pp. 84–-98, 2017.

\bibitem{ref34}
S. H. Chan, ``Performance analysis of Plug-and-Play ADMM: A graph signal processing perspective,'' \textit{IEEE Trans. Comput. Imaging}, vol. 5, no. 2, pp. 274-286, 2019.

\bibitem{ref35}
S. Sreehari et al., ``Plug-and-Play priors for bright field electron tomography and sparse interpolation,'' \textit{IEEE Trans. Comput. Imaging}, vol. 2, no. 4, pp. 408–423, 2016.

\bibitem{ref36}
A. M. Teodoro, J. M. Bioucas-Dias, M. A. T. Figueiredo, ``Scene-adapted Plug-and-Play algorithm with convergence guarantees,'' in \textit{Proc. IEEE 27th Int. Workshop Mach. Learn. Signal Process. (MLSP)}, Tokyo, Japan, 2017, pp. 1--6.

\bibitem{ref37}
Y. Sun, Z. Wu, X. Xu, B. Wohlberg and U. S. Kamilov, ``Scalable Plug-and-Play ADMM with convergence guarantees,'' \textit{IEEE Trans. Comput. Imaging}, vol. 7, pp. 849--863, 2021.

\bibitem{ref38}
K. Wei, A. Aviles-Rivero, J. Liang, et al., ``TFPNP: Tuning-free plug-and-play proximal algorithms with applications to inverse imaging problems,'' \textit{J. Mach. Learn. Res.}, vol. 23, no. 16, pp. 1-–48, 2022.

\bibitem{ref39}
Y. Wang, Z. He, X. Zhan, Q. Zeng and Y. Hu, ``A 3-D sparse SAR imaging method based on Plug-and-Play,'' \textit{IEEE Trans. Geosci. Remote Sens.}, vol. 60, pp. 1--14, 2022.

\bibitem{ref40}
Y. Romano, M. Elad, P. Milanfar, ``The little engine that could: Regularization by Denoising (RED),'' \textit{Siam J. Imaging Sci.}, vol. 10, no. 4, pp. 1804--1844, 2017.

\bibitem{ref41}
E. T. Reehorst and P. Schniter, ``Regularization by Denoising: Clarifications and new interpretations,'' \textit{IEEE Trans. Comput. Imaging}, vol. 5, no. 1, pp. 52--67, 2019.

\bibitem{ref42}
Y. Wang, X. Zhan, J. Yao, et al., ``3D sparse SAR imaging based on complex-valued nonconvex regularization for scattering diagnosis,'' \textit{IEEE Antennas Wirel. Propag. Lett.}, vol. 23, no. 2, pp. 888--892, 2024.

\bibitem{ref43}
Z. Xu, B. Zhang, G. Zhou, et al., ``Sparse SAR imaging and quantitative evaluation based on nonconvex and TV regularization,'' \textit{Remote Sens.}, vol. 13, no.9, pp. 1643, 2021.

\bibitem{ref44}
A. Güngör, M. Çetin, H. E. Güven, ``Compressive synthetic aperture radar imaging and autofocusing by augmented Lagrangian methods,'' \textit{IEEE Trans. Comput. Imaging}, vol. 8, pp. 273--285, 2022.

\bibitem{refs2}
J. Gao, Y. Wang, J. Yao, X. Zhan, G. Sun and J. Bai, ``Three-dimensional array SAR sparse imaging based on hybrid regularization,'' \textit{IEEE Sensors J.}, 2024, doi: 10.1109/JSEN.2024.3386901.

\bibitem{ref45}
S. Ono, ``Primal-dual Plug-and-Play image restoration,'' \textit{IEEE Signal Process. Lett.}, vol. 24, no. 8, pp. 1108--1112, 2017.

\bibitem{ref46}
Z. Xu, B. Zhang, Z. Zhang, et al., ``Nonconvex-nonlocal total variation regularization-based joint feature-enhanced sparse SAR imaging,'' \textit{IEEE Geosci. Remote Sens. Lett.}, vol. 19, pp. 1--5, 2022.

\bibitem{ref47}
M. V. Afonso, J. M. Bioucas-Dias and M. A. T. Figueiredo, ``Fast image recovery using variable splitting and constrained optimization,'' \textit{IEEE Trans. Image Process.}, vol. 19, no. 9, pp. 2345--2356, 2010.

\bibitem{ref48}
J. Nocedal and S. J. Wright, {\it{Numerical optimization}}, 2nd ed. New York: Springer-Verlag, 2006.

\bibitem{ref49}
Z. Wei, B. Zhang, Z. Xu, B. Han, W. Hong, and Y. Wu, ``An improved SAR imaging method based on nonconvex regularization and convex optimization,'' \textit{IEEE Geosci. Remote Sens. Lett.}, vol. 16, no. 10, pp. 1580–-1584, 2019.

\bibitem{refz2}
{Y. Liu, X. Yuan, J. Suo, D. J. Brady and Q. Dai, ``Rank Minimization for Snapshot Compressive Imaging,'' \textit{IEEE Trans Pattern Anal Mach Intell.}, vol. 41, no. 12, pp. 2990–-3006, 2019.}

\bibitem{refz3}
{R. T. Rockafellar and R. J.-B. Wets, ``Variational analysis,'' \textit{Springer Science \& Business Media}, vol. 317, 2009.}

\bibitem{refz4}
{H. H. Bauschke, S. M. Moffat, and X. Wang, ``Firmly nonexpansive mappings and maximally monotone operators: correspondence and duality,'' \textit{Set-Valued and Variational Analysis}, vol. 20, pp. 131, 2012.}

\bibitem{refzz1}
{R. Cohen, M. Elad, P. Milanfar, ``Regularization by denoising via fixed-point projection (RED-PRO),'' \textit{SIAM J. Imag. Sci.}, vol. 14, pp. 1374--1406, 2021.}

\bibitem{refz5}
{S. Bartz, H. H. Bauschke, J. M. Borwein, S. Reich, and X. Wang, ``Fitzpatrick functions, cyclic monotonicity and Rockafellar’s antiderivative,'' \textit{Nonlinear Analysis: Theory, Methods \& Applications}, vol. 66, pp. 1198--1223, 2007.}

\bibitem{ref52}
Z. Wei, B. Zhang, Z. Xu, B. Han, W. Hong, and Y. Wu, ``Edge preserved low-rank SAR image despeckling via hierarchical prior knowledge regulation,'' \textit{IEEE Trans. Geosci. Remote Sens.}, vol. 61, pp. 1--17, 2023.

\bibitem{refz6}
{K. Zhang, W. Zuo, S. Gu and L. Zhang, ``Learning deep CNN denoiser prior for image restoration,'' in \textit{Proc. IEEE Conf. Comput. Vis. Pattern Recognit.}, Honolulu, HI, USA, 2017, pp. 2808--2817.}

\end{thebibliography}
\end{document}